\documentclass[english]{revtex4}
\usepackage[T1]{fontenc}
\usepackage[latin9]{inputenc}
\usepackage{color}
\usepackage{amsmath}
\usepackage{amssymb}
\usepackage{graphicx}
\usepackage{babel}
\usepackage{mathrsfs}

\begin{document}

\title{On thermodynamic inconsistencies in several photosynthetic and solar cell models and how to fix them}

\author{David Gelbwaser-Klimovsky and Al\'an Aspuru-Guzik}

\affiliation{Department of Chemistry and Chemical Biology, Harvard University,
Cambridge, MA 02138}

\begin{abstract}
We analyze standard theoretical models of solar energy conversion developed to study solar
cells and  photosynthetic systems. We show that the assumption
that the energy transfer to the reaction center/electric circuit is
through a decay rate or ``sink'', is in contradiction with the second
law of thermodynamics. We put forward a thermodynamically consistent
alternative by explicitly considering parts of the reaction center/electric
circuit and by employing a Hamiltonian transfer. The predicted energy
transfer by the new scheme differs from the one found using a decay
rate, casting doubts on the validity of the conclusions obtained by
models which include the latter.
\end{abstract}
\maketitle

Light-harvesting organism and solar cells convert thermal photons from the sun,
into  useful energy such as ATP or electric
power \cite{blankenship2013molecular,nelson2003physics,wurfel2009physics}.
Understanding and improving these processes may led to more efficient ways to produce clean energy (see \cite{blankenship2011comparing}
and references within).
These systems are effectively heat engines
\cite{alicki2015solar,einax2014network,kondepudi2014modern,gelbwaser2015thermodynamics} because they transform
a heat flow into power (useful energy). Therefore, they are constrained
by the laws of thermodynamics \cite{shockley1961detailed,blankenship2011comparing,landsberg1980thermodynamic,knox2007entropy,knox1969thermodynamics,parson1978thermodynamics,ross1967thermodynamics,alicki2015non}
which set a fundamental efficiency bound based on the distinction
between the two forms of energy exchange: heat flow and power. These
two are not interchangeable: in a cyclic process, power may be totally
converted into heat flow, but the opposite is forbidden by the second
law of thermodynamics \cite{kondepudi2014modern,gelbwaser2013work,gelbwaser2014heat}.

A key for understanding the efficiency and the power produced by  solar cells and
plants, is the development of microscopical models of the
energy absorption, transmission and storage. Previous works have proposed
that effects such environment assisted quantum transport \cite{mohseni2008environment,rebentrost2009environment,plenio2008dephasing},
coherent nuclear motion \cite{novoderezhkin2004coherent,killoran2014enhancing},
as well as quantum coherences \cite{dorfman2013photosynthetic,scully2011quantum,creatore2013efficient,fassioli2010quantum},
play an important role in the enhancement of the energy conversion.

For practical computational and theoretical reasons, models have been restricted to the study of specific subsystems. It is customary to study photosynthetic complexes coupled to "traps" or "sinks" that represent the reaction center where exciton dissociation occurs \cite{mohseni2008environment,plenio2008dephasing,rebentrost2009environment}. Similar models have been employed for the study of exciton absorption and transport in, e.g., organic solar cells \cite{alharbi2015theoretical,cao2009optimization,creatore2013efficient,dorfman2013photosynthetic,killoran2014enhancing,scully2011quantum}.

Here we show that if not careful, the introduction of sinks and traps leads to violations of the second law of thermodynamics. These violations are a reason of concern for the validity of the models that have been employed to date. To shed light on the issue and to provide a simple to understand situation, we introduce 
a toy model to study this approximation and put forward a thermodynamically
consistent version of it. This model could be used as the basis for more elaborate solar
cell and plant microscopic models. Finally, we show that the output power of the thermodynamically-consistent version of the model can differ substantially to the simple trap or sink models.

\subsection*{Second law of thermodynamics}

\begin{figure*}
\begin{centering}
\includegraphics[width=1\textwidth]{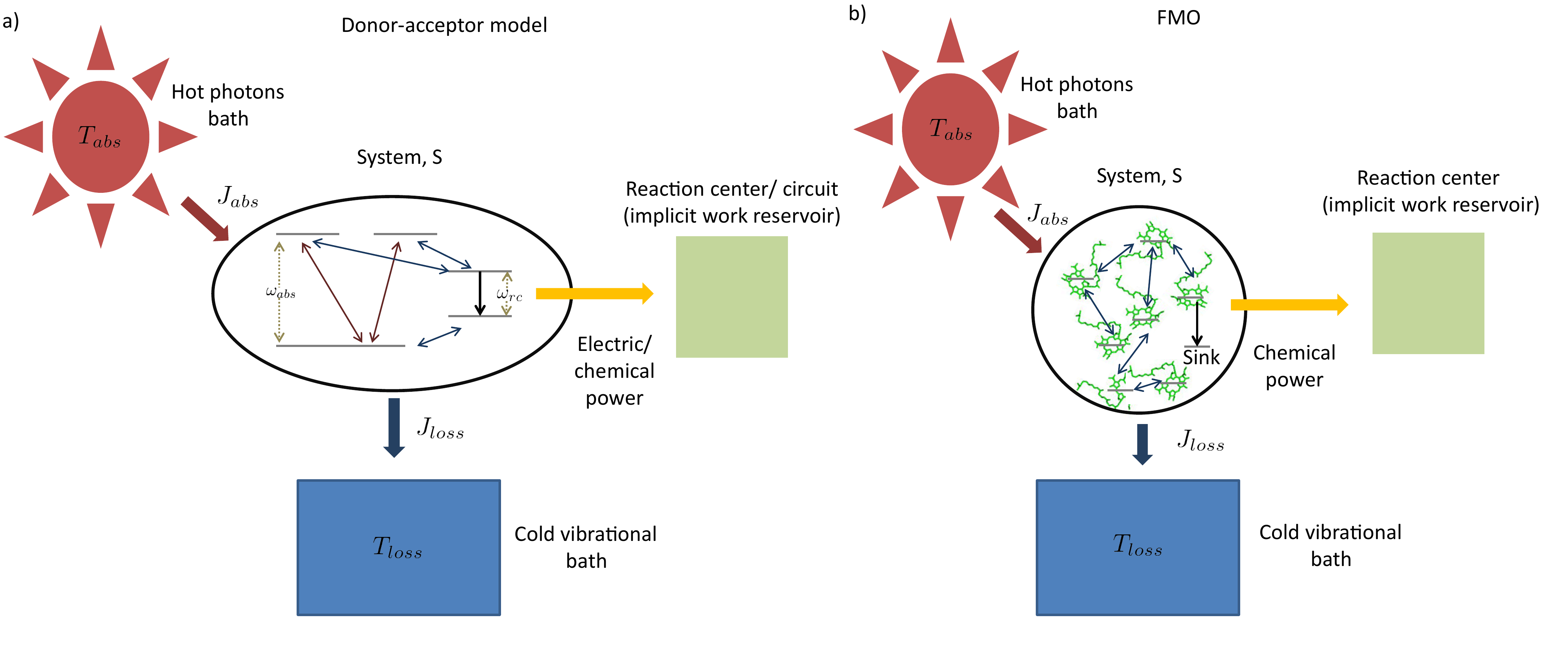}
\par\end{centering}
\caption{(Color online) Solar energy conversion models: a) donor-acceptor model; b) FMO model.
In both cases the allowed transitions are shown only  for illustration purposes
and may vary between different models. }
\label{fig:sketch}
\end{figure*}

The standard thermodynamic models for solar energy conversion are comprised by a system,
S, that interacts with different thermal baths and transforms the
solar energy into chemical energy or electric current. Here we analyze two types
of models: donor-acceptor models, where S is composed of four to five levels. These models have been applied for studying solar cells \cite{creatore2013efficient,scully2011quantum}
as well as  photosynthetic systems \cite{dorfman2013photosynthetic}
(see Figure \ref{fig:sketch}a); or models of the celebrated Fenna-Matthews-Olson (FMO) complex models, where S includes
seven bacteriochlorophyll, each of them  described by a single energy
state \cite{alharbi2015theoretical,cao2009optimization,killoran2014enhancing,mohseni2008environment,plenio2008dephasing,rebentrost2009environment}
(see Figure \ref{fig:sketch}b). In both cases, the energy conversion
process is composed of the following explicit or implicit steps: i) \textit{Light absorption}. The system,
S, absorbs hot photons coming from the sun. The temperature of the
photon is $T_{abs}$ and $J_{abs}$ is the heat flow between the hot
photons and S; ii) \textit{Energy transfer}. The absorbed
energy is transmitted between different states of the  S. The
number of states and allowed transitions varies from case to case.
During this stage some energy is lost through a heat current,
$J_{loss}$, to a vibrational bath  at room temperature $T_{loss}$ (material photons for solar cells
\cite{wurfel2009physics,nelson2003physics} or protein modes for photosynthetic
systems \cite{blankenship2013molecular}); iii) \textit{Power extraction}. A decay rate that represents an
irreversible energy flow to an external system, work reservoir.
 The latter is generally not
explicitly considered. For photosynthetic models this last stage,
involves the decay to a sink or trap, together with an energy transfer to the RC and its subsequent transformation  into
chemical energy. In the case of solar cells, the energy flow is the electric power
that runs through the circuit.  

The dynamics of these systems is constrained by the second law of thermodynamics, through the  entropy production inequality
 \cite{spohn1978entropy,gelbwaser2015thermodynamics},

\begin{equation}
\sigma=\dot{S}(\rho_{s})-\frac{J_{abs}}{T_{abs}}-\frac{J_{loss}}{T_{loss}}\geq0,\label{eq:secondlaw}
\end{equation}

\noindent where $\sigma$ is the entropy production, $\rho_{s}$ is S
density matrix and $\dot{S}$ is the derivative over time of the Von-Neumann
entropy \cite{breuer2002theory}. For the heat currents, as wells as for the power, we use the sign convention
that energy flowing to and from   S is positive and negative respectively.   Models with artificial sinks could be envisioned as systems that transfer energy to a zero-temperature bath. This will justified the addition of an extra term on the r.h.s of Eq. \eqref{eq:secondlaw}.  In such circumstances the efficiency of the system, in principle can be up to 100\%.
Nevertheless, solar cells and plants must obey the same thermodynamic bound as a  heat engine operating between thermal baths at the temperatures of the sun and the vibrational bath, which are 6000k and 300k respectively and therefore bounded to  $95\%$. This is a maximum absolute bound based solely on the temperatures. In more elaborate models, the bound is even lower \cite{shockley1961detailed,blankenship2011comparing,landsberg1980thermodynamic,knox2007entropy,knox1969thermodynamics,parson1978thermodynamics,ross1967thermodynamics,alicki2015non}.

In the case of a steady state flux of solar energy into S, the state of S on
average does not change, and the second law, Eq. \ref{eq:secondlaw},
simplifies to 

\begin{gather}
\frac{-J_{loss}}{J_{abs}}\geq\frac{T_{loss}}{T_{abs}},\quad J_{abs}>0\nonumber, \\
\frac{-J_{loss}}{J_{abs}}\leq\frac{T_{loss}}{T_{abs}},\quad J_{abs}<0.\label{eq:sssecond}
\end{gather}

The donor/acceptor models studied in \cite{dorfman2013photosynthetic,scully2011quantum,creatore2013efficient,fassioli2010quantum},
analyze the solar energy conversion at steady state, and their heat
currents ratio has the form (see SI):

\begin{gather}
\frac{-J_{loss}}{J_{abs}}=1-\frac{\omega_{rc}}{\omega_{abs}},\label{eq:heatcurrates}
\end{gather}

\noindent where $\omega_{abs}$ is the energy of the absorbed photons and $\omega_{rc}$
is the energy of the excitation transferred to the RC/circuit (work
reservoir) (see Figure \ref{fig:sketch}a).  In
all these models, the signs of the currents are independent of the parameters,
$J_{loss}<0$ and $J_{abs}>0$ (see SI). 

\begin{figure*}
\centering{}
\includegraphics[width=1\textwidth]{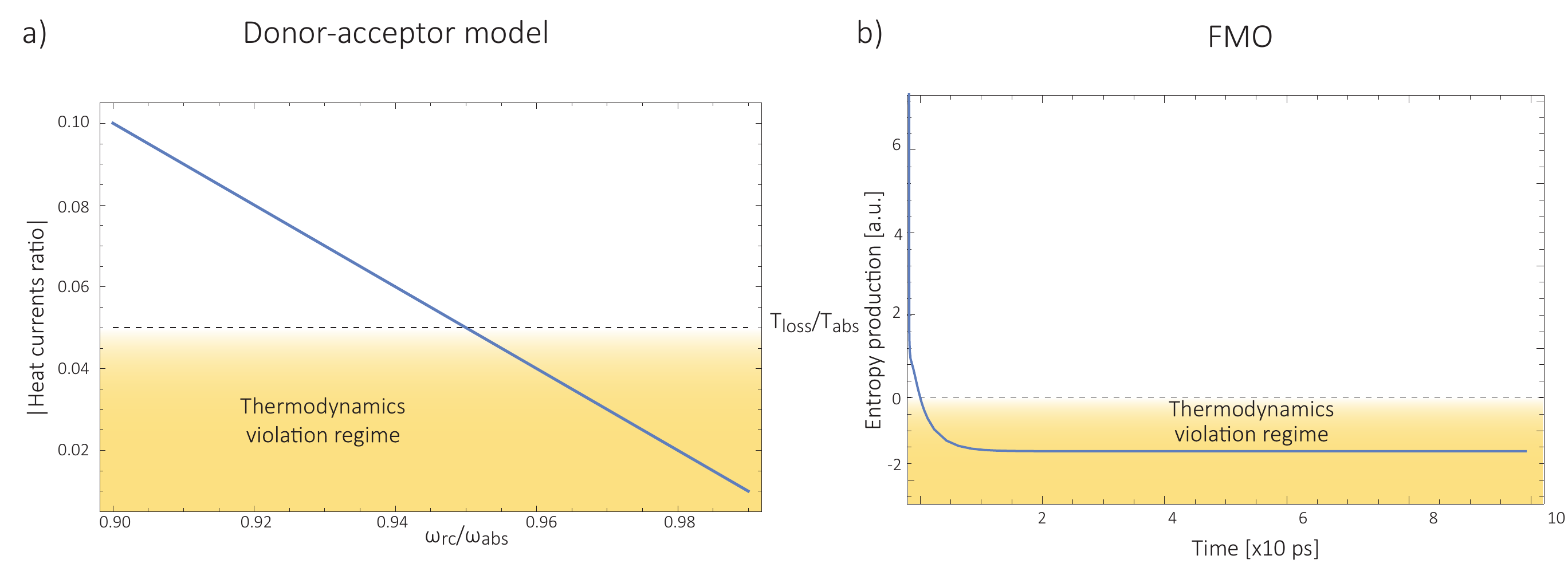}\caption{(Color online) Violation of thermodynamics by models of solar energy conversion.
a)  Absolute value of the heat currents ratio  as function of frequency ratio for the steady
state models on references \cite{dorfman2013photosynthetic,scully2011quantum,creatore2013efficient,fassioli2010quantum}.
For  large $\omega_{rc}$, see Eqs. \eqref{eq:sssecond} and \eqref{eq:heatcurrates}, these models break the second
law of thermodynamics; b) Entropy production as
a function of time for FMO models \cite{alharbi2015theoretical,cao2009optimization,killoran2014enhancing,mohseni2008environment,plenio2008dephasing,rebentrost2009environment,caruso2009highly}.
In both graphs the shaded area represents a regime forbidden by thermodynamics.}\label{fig:violth}
\end{figure*}

As shown in Figure \ref{fig:violth}a, for $1-\frac{T_{loss}}{T_{abs}}\leq\frac{\omega_{rc}}{\omega_{abs}},$
these models violate the second law of thermodynamics. Realistic model parameters may well fall outside of this range. This does not exclude the fact that the model is both inconsistent and potentially leading to artificial results. As we
show below, the power predicted by a thermodynamically consistent
model differs from the simple sink or trap models.

Figure \ref{fig:violth}b shows the entropy production (Eq.\ref{eq:secondlaw})
as function of time for standard sink or trap models of the FMO  complex \cite{alharbi2015theoretical,cao2009optimization,killoran2014enhancing,mohseni2008environment,plenio2008dephasing,rebentrost2009environment,caruso2009highly}. A simplified model is used for the antenna (a
two level system), which is coupled to the FMO. The energy is transferred
to the RC (work reservoir) through a decay term (see Figure \ref{fig:sketch}b
and SI). In this scenario the dynamics outside the steady state is
considered. For these models, there is not a simple analytical expression such as Eq. \ref{eq:heatcurrates},
therefore we use a standard numeric simulation based on a Lindblad
equation \cite{davies1974markovian,gorini1976completely,lindblad1976generators}.
As seen in Figure \ref{fig:violth}b, these models also violate the
second law of thermodynamics. Details of our model can be found in the SI.

\subsection*{Thermodynamically-consistent model}

The assumption in the trap or sink models  that the  energy transfer to the RC/circuit is based solely
on a relaxation process, introduces an inconsistency with thermodynamics.
Even though physically this energy flow is power, a decay
rate effectively represents a heat flow to a thermal bath. This is
the root of the inconsistency. Here we use a toy model to clarify
this point and put forward an alternative that could serve as basis
to correctly model these systems. We compare between two possible
energy transfers schemes to the RC/circuit: i) standard decay;
ii)  Hamiltonian transfer. 

\begin{figure*}

\begin{centering}
\includegraphics[width=1\textwidth]{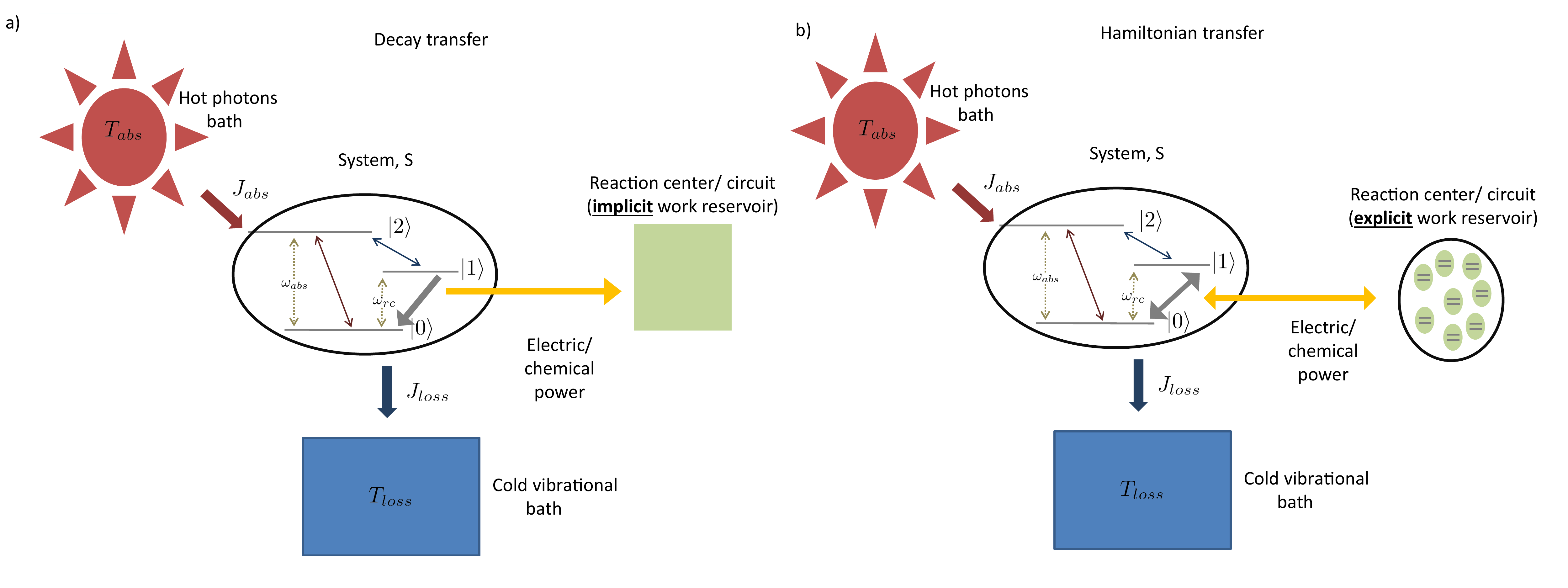}
\par\end{centering}
\caption{(Color online) A toy model used to study different energy transfer schemes: decay
rate (left); Hamiltonian transfer (right).}\label{fig:toy models}
\end{figure*}

As S, we consider a three level system as shown in Figure \ref{fig:toy models}.
The absorption of a photon  causes an  excitation transfer between $|0\rangle$
and $|2\rangle$, whereas phonons are emitted by  transitions from
$|2\rangle$ to $|1\rangle$. Finally, the cycle is closed by a transition
between $|1\rangle$ and $|0\rangle$, and the energy difference is
transferred to the RC/circuit. 

For both schemes  the S-bath Hamiltonian is 

\begin{equation}
H_{s}+H_{B}+H_{SB}.
\label{eq:hamtot}
\end{equation}

The S Hamiltonian, in natural units ($\hbar=1$ and $k_{B}=1$) is

\begin{eqnarray}
H_{S} & = & \omega_{abs}|2\rangle\langle2|+\frac{\omega_{rc}}{2}\left(|1\rangle\langle1|-|0\rangle\langle0|\right)
\end{eqnarray}

\noindent and $H_{B}=H_{Photons}+H_{Phonons}$ are the photon  and phonon bath free Hamiltonian.
Both baths are in thermal equilibrium at temperatures $T_{abs}$ and
$T_{loss},$ respectively. The S-bath interaction is governed
by 
\begin{gather}
H_{SB}= 
\sum_{\lambda}g_{h,\lambda}\left(|2\rangle\langle0|a_{\lambda}+|0\rangle\langle2|a_{\lambda}^{\dagger}\right) \nonumber\\
+\sum_{\lambda}g_{c,\lambda}\left(|2\rangle\langle1|b_{\lambda}+|1\rangle\langle2|b_{\lambda}^{\dagger}\right),
\end{gather}

\noindent where $a_{\lambda},$$a_{\lambda}^{\dagger}$ ($b_{\lambda},$$b_{\lambda}^{\dagger}$)
are the annihilation and creation operator of photons (phonons) modes.
We assume that the baths are Markovian and are weakly coupled to S \cite{breuer2002theory}. For the sake of simplicity, we assume
that the zero temperature decay rates \cite{gordon2009cooling} of both baths are the same as the transfer rate to the RC/circuit, $\Gamma_h=\Gamma_c=\Gamma$ (see SI). 

 \textit{i) Decay transfer}

The standard relaxation scheme is a decay rate between $|1\rangle$
and $|0\rangle$,

\begin{equation}
H_{trans}^{Dec}=\sqrt{\Gamma}|0\rangle\langle1|,
\label{eq:Htd}
\end{equation}

\noindent where the RC/circuit is not explicitly included;  

 \textit{ii) Hamiltonian transfer}

An alternative to the model above is to explicitly include at least
part of the RC/circuit, which plays the role of the work reservoir.
In photosynthetic systems, the last stage on the reaction center is
the transfer of electrons to the $Q_{B}$ quinone, that once is full,
migrates to further proceed with the ATP production \cite{jones2009petite}.
This quinone is replaced by an empty one  from a quinone pool. Inspired
by this process, we construct a toy model of the work reservoir that
could be a guideline for more complicated photosynthetic or solar cells
models. It consists of a collection of independent and
identical two level systems (TLS). Each of them represents a quinone in a
photosynthetic system or an electrode site in a solar cell. The ground state corresponds to an empty quinone/site, and the excited state to a ``full'' quinone/site. Furthermore, we assume that there are always empty quinones/sites
available to accept an  electron. Thus, the number of quinones/sites,
$j,$ is always much larger than the number of electrons $c^{\dagger}c$, $j\gg c^{\dagger}c$. This assumption is equivalent to
the thermodynamic limit taken in the Holstein-Primakoff procedure
\cite{holstein1940field,emary2004phase}, which allows to describe
the collection of quinones/sites as a single harmonic oscillator (HO).
Therefore, we can write the  work reservoir and transfer Hamiltonian
as (see SI)

\begin{gather}
H_{trans}^{Ham}=\sqrt{\Gamma}\left(c|1\rangle\langle0|+c^{\dagger}|0\rangle\langle1|\right)+
\omega_{rc}(c^{\dagger}c-j),
\end{gather}

\noindent where $c$, $c^{\dagger}$ are the annihilation and creation operator
of the HO. Furthermore,
for the sake of simplicity we assume that the HO is resonant with
the $|1\rangle\leftrightarrow|0\rangle$ transition and that  is weakly coupled to S, $\omega_{rc}\gg\Gamma$.

In order to find the energy that is being transferred, in 
both schemes  we first solve the dynamic equations. For this we use 
the standard Born-Markov approximation \cite{breuer2002theory}  and
write the Lindblad equations for (see SI): \textit{i)} the three level system
in the case of the decay rate scheme; \textit{ii)} the three level system and the
HO for the Hamiltonian transfer scheme, which are at product state due to the weak coupling between them. For both schemes, we analyze the energy transfer
at the three level system  steady state. 

\textit{i)} For the decay transfer the excitations rate  to the RC/circuit is $\Gamma\rho^{ss}_{11},$ and the power is (see SI)

\begin{gather}
P^{Dec}=-\omega_{rc}\Gamma\rho_{11}^{ss},\nonumber \\
\rho_{11}^{ss}= 
\frac{1}{1+2e^{\left(\omega_{abs}+\frac{\omega_{rc}}{2}\right)/T_{abs}}},\label{eq:pdec}
\end{gather}

\noindent where $\rho_{11}^{ss}$ is the steady state population of level $|1\rangle$.
Power is always extracted ($P^{Dec}<0)$, even if the temperatures
are the same, $T_{abs}=T_{loss}$. This is in  contradiction with thermodynamics,
which forbids cyclic power extraction in the presence of a single
temperature. A further evidence of the  violation of thermodynamics is the combination between the temperature independence of the heat currents ratio and the positivity of $J_{abs}^{Dec}$ (Eqs. \eqref{eq:sssecond} and \eqref{eq:heatcurrates}),

\begin{gather}
\frac{-J_{loss}^{Dec}}{J_{abs}^{Dec}}=
\frac{2\omega_{abs}-\omega_{rc}}{2\omega_{abs}+\omega_{rc}}, \nonumber \\
 J_{abs}^{Dec}=\left(\omega_{abs}+\frac{\omega_{rc}}{2}\right)\Gamma\rho_{11}^{ss}>0.
\end{gather}

For $\frac{2\omega_{abs}-\omega_{rc}}{2\omega_{abs}+\omega_{rc}}<\frac{T_{loss}}{T_{abs}}$
the model breaks the second law of thermodynamics, Eq. \eqref{eq:sssecond}. 

\textit{ii)} The  power extraction for the Hamiltonian transfer
differs from $P^{Dec}$ (see SI),

\begin{equation}
P^{Ham}=-\omega_{rc}\langle\dot{n}\rangle=-\omega_{rc}(s-r),
\label{eq:ph}
\end{equation}

\begin{figure*}
\begin{centering}
\includegraphics[width=1\textwidth]{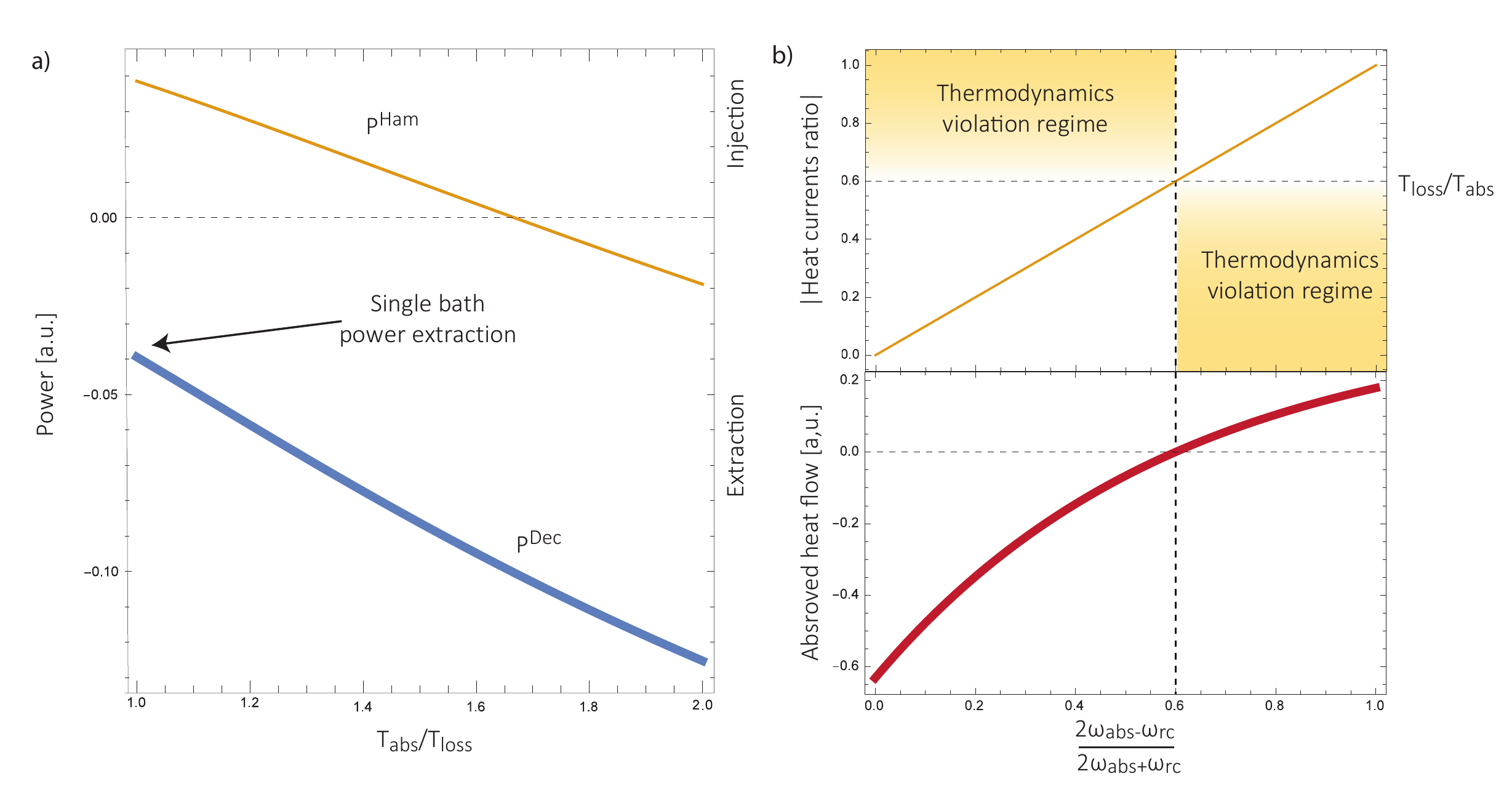}
\par\end{centering}
\caption{(Color online) a) Predicted power extraction for the decay  ($P^{Dec}$,
thick blue line) and the Hamiltonian  ($P^{Ham},$ thin yellow
line) transfer schemes. The former  predicts power extraction, $P^{Dec}<0$,
from a single bath ($\frac{T_{loss}}{T_{abs}}=1$), while the latter
does not ($P^{Ham}>0$). b)  Absolute value of the heat currents ratio    (yellow thin line)
for the Hamiltonian transfer scheme (top). The sign change of $J_{abs}^{Ham},$
(bottom, thick red line) splits the regions forbidden by thermodynamics (shaded
areas), preventing its violation. In contrast,  for the  decay transfer scheme, $J_{abs}^{Dec}$ is always positive, preventing the thermodynamically forbidden region splitting and placing the heat currents ratio in a thermodynamically forbidden region (see Figure \ref{fig:violth}a).}\label{fig:Powers}
\end{figure*}	

\noindent  $\langle\dot{n}\rangle$ is the HO population change. We have
assumed an ideal case, where all the energy flow to the HO is considered
as power, which just represents a  maximum bound \cite{gelbwaser2013work,gelbwaser2014heat}.
The heat currents are (see SI)

\begin{gather}
J_{abs}^{Ham}=\left(\omega_{abs}+\frac{\omega_{rc}}{2}\right)(s-r), \nonumber\\
J_{loss}^{Ham}=-\left(\omega_{abs}-\frac{\omega_{rc}}{2}\right)(s-r)
\end{gather}

\noindent and 

\begin{gather}
s-r= \nonumber \\
K_{1}\left(e^{-\left(\omega_{abs}+\frac{\omega_{rc}}{2}\right)/T_{abs}}-e^{-\left(\omega_{abs}-\frac{\omega_{rc}}{2}\right)/T_{loss}}\right),
\end{gather}

\noindent where $K_{1}$ is always positive and depends on the couplings to baths
(see SI). In contrast to  the decay transfer scheme, in this case
power is extracted, $P^{Ham}<0$, only for certain combination of
parameters, 

\begin{gather}
\frac{T_{loss}}{T_{abs}}<\frac{2\omega_{abs}-\omega_{rc}}{2\omega_{abs}+\omega_{rc}}
\end{gather}

\noindent and power can not be extracted if both temperatures are the same. Further
divergences between $P^{Dec}$ and $P^{Ham}$ can be seen in Figure
\ref{fig:Powers}a.

Figure \ref{fig:Powers}b shows that the  heat currents ratio of the Hamiltonian
transfer scheme complies with the second law of thermodynamics (see
Eq \ref{eq:sssecond}). The thermodynamic violation regime  splits
due to $J_{abs}^{Ham}$ sign change. Although for positive $J_{abs}^{Ham}$,
the absolute value of the heat currents ratio   should be larger than the temperatures ratio,
for negative $J_{abs}^{Ham}$, it should be smaller. The lack of sign
change for $J_{abs}^{Dec}$, prevents the splitting of the thermodynamic
violation regime, placing the heat currents ratio in a thermodynamically forbidden region (see Figure \ref{fig:violth}a).

\subsection*{Conclusions}

We have analyzed several models used for describing energy absorption
and transmission both in solar cells and in photosynthetic systems
such as the FMO complex. We have shown that the use of sinks, traps or any artificial relaxation
process in order to describe the energy transfer to a further stage
(the reaction center in photosynthetic systems  or the  electric circuit in a solar cell) introduces a contradiction with the second law of thermodynamics.
This invalidates several models currently used to study solar energy conversion,  casting doubts regarding   their conclusions. These includes the role of coherences, environment assisted quantum transport, coherent nuclear motion and the presence
of quantum effects in photosynthesis, among others. We do not argue against the
existence of those effects in the conversion of solar energy. But they
should be verified using thermodynamically consistent models.

We have further proposed how to correctly analyze these systems. We
show this in a thermodynamically consistent toy model that explicitly
describes parts of the RC/circuit and uses a Hamiltonian term to describe
the energy transfer instead of a decay rate. The predicted transmitted energy
 greatly differs between these two alternatives (see Figure
\ref{fig:Powers}a), highlighting the  need to review the conclusions
derived by thermodynamically inconsistent models.

\section*{Acknowledgments}

 We acknowledge Robert Alicki and Doran Bennett for useful discussions. We acknowledge the support from the Center for Excitonics, an Energy
Frontier Research Center funded by the U.S. Department of Energy under award DE-SC0001088 (Solar energy conversion process). D. G-K. also acknowledges the support of the CONACYT (Quantum thermodynamics).


\section*{Supplementary information}
\setcounter{equation}{0}
\renewcommand{\theequation}{S\arabic{equation}}

\setcounter{figure}{0}
\renewcommand{\thefigure}{S\arabic{figure}}
\section{Energy conversion models}

We derive the evolution equations for some examples of two types of
energy conversion models. The results of this section are used to
generate Figure 2 in the main text, as well as Eq. 3. Unless otherwise
stated, we assume $\hbar=k_{b}=1$.

\subsection{Donor-acceptor models}

As examples of these models, we analyze below two particular donor-acceptor
models that use a decay transfer scheme. This kind of analysis may
be expanded to models that include coherent vibronic evolution such
as the proposed on \cite{killoran2014enhancing}.

1) We consider the biological quantum heat engine model proposed on
\cite{dorfman2013photosynthetic} (see in particular Eqs. S34-S37 on \cite{dorfman2013photosynthetic})
. It consists of a four level system coupled to a hot bath, a cold
bath, and to the reaction center/circuit (also termed ``the load'').
$T_{h(c)}$ is the hot (cold) bath temperature. The different decay
rates are shown in Figure S1. The equations of motion are

\begin{gather}
\dot{\rho}_{aa}=-\gamma_{c}\left[\left(1+\bar{n}_{c}\right)\rho_{aa}-\bar{n}_{c}\rho_{\alpha\alpha}\right]-\gamma_{h}\left[\left(1+\bar{n}_{h}\right)\rho_{aa}-\bar{n}_{h}\rho_{bb}\right],\nonumber \\
\dot{\rho}_{\alpha\alpha}=\gamma_{c}\left[\left(1+\bar{n}_{c}\right)\rho_{aa}-\bar{n}_{c}\rho_{\alpha\alpha}\right]-\Gamma\rho_{\alpha\alpha},\nonumber \\
\dot{\rho}_{bb}=\gamma_{h}\left[\left(1+\bar{n}_{h}\right)\rho_{aa}-\bar{n}_{h}\rho_{bb}\right]+\Gamma_{c}\left[\left(1+\bar{N}_{c}\right)\rho_{\beta\beta}-\bar{N}_{c}\rho_{bb}\right],\nonumber \\
\rho_{aa}+\rho_{bb}+\rho_{\alpha\alpha}+\rho_{\beta\beta}=1,\label{eq:meq1}
\end{gather}
where we have kept the original paper notation. $\rho_{ii}$ is the
level population of state $i$ and $\bar{n}_{i}$ or $\bar{N}_{i}$
are the relevant i- bath mode population. For details on Eq. \ref{eq:meq1}
derivation, we refer the reader to the original paper.  The steady
state populations are 

\begin{gather}
\frac{\rho_{aa}^{ss}}{\rho_{\alpha\alpha}^{ss}}=\frac{\gamma_{c}\bar{n}_{c}+\Gamma}{\gamma_{c}(\bar{n}_{c}+1)},\\
\frac{\rho_{bb}^{ss}}{\rho_{\alpha\alpha}^{ss}}=\frac{\Gamma\left[\gamma_{c}\left(\bar{n}_{c}+1\right)+\gamma_{h}\left(\bar{n}_{h}+1\right)\right]+\gamma_{h}\gamma_{c}\bar{n}_{c}\left(1+\bar{n}_{h}\right)}{\gamma_{h}\bar{n}_{h}\gamma_{c}\left(\bar{n}_{c}+1\right)},\\
\frac{\rho_{aa}^{ss}}{\rho_{bb}^{ss}}=\frac{\left(\gamma_{c}\bar{n}_{c}+\Gamma\right)\gamma_{h}\bar{n}_{h}}{\Gamma\left\{ \gamma_{c}\left(\bar{n}_{c}+1\right)+\gamma_{h}\left(\bar{n}_{h}+1\right)\right\} +\gamma_{h}\gamma_{c}\bar{n}_{c}\left(1+\bar{n}_{h}\right)},\\
\frac{\rho_{\beta\beta}^{ss}}{\rho_{bb}^{ss}}=e^{-(\omega_{\beta}-\omega_{b})/T_{c}}+\frac{\gamma_{h}\bar{n}_{h}\Gamma\gamma_{c}\left(\bar{n}_{c}+1\right)}{\left(1+\bar{N}_{c}\right)\Gamma_{c}\left[\Gamma\left\{ \gamma_{c}\left(\bar{n}_{c}+1\right)+\gamma_{h}\left(\bar{n}_{h}+1\right)\right\} +\gamma_{h}\gamma_{c}\bar{n}_{c}\left(1+\bar{n}_{h}\right)\right]}.
\end{gather}

The heat currents are defined as the energy flow between the four
level system and the i-bath,

\begin{gather}
J_{i}=Tr[\mathcal{L}_{i}\left(\rho\right)H_{S}],\label{eq:heatcurrents}
\end{gather}
where $\mathcal{L}_{i}\left(\rho\right)$ is the reduced evolution
induced only by the i-bath and $H_{S}$ is the four level Hamiltonian.
The heat currents at steady state are 

\begin{gather}
J_{h}=(\omega_{a}-\omega_{b})\left(1+\bar{n}_{h}\right)\gamma_{h}\rho_{bb}^{ss}\left(e^{-(\omega_{a}-\omega_{b})/T_{h}}-\frac{\rho_{aa}^{ss}}{\rho_{bb}^{ss}}\right)=\nonumber \\
\frac{\left(\omega_{a}-\omega_{b}\right)\left(1+\bar{n}_{h}\right)\gamma_{h}\rho_{bb}^{ss}}{\Gamma\left[\gamma_{c}\left(\bar{n}_{c}+1\right)+\gamma_{h}\left(\bar{n}_{h}+1\right)\right]+\gamma_{h}\gamma_{c}\bar{n}_{c}\left(1+\bar{n}_{h}\right)}\left(e^{-(\omega_{a}-\omega_{b})/T_{h}}\Gamma\gamma_{c}(\bar{n}_{c}+1)\right),\\
J_{c}=\left(\omega_{a}-\omega_{\alpha}\right)\left(1+\bar{n}_{c}\right)\gamma_{c}\rho_{\alpha\alpha}^{ss}\left(e^{-(\omega_{a}-\omega_{\alpha})/T_{c}}-\frac{\rho_{aa}^{ss}}{\rho_{\alpha\alpha}^{ss}}\right)+\left(\omega_{\beta}-\omega_{b}\right)\left(1+\bar{N}_{c}\right)\Gamma_{c}\rho_{bb}^{ss}\left(e^{-(\omega_{\beta}-\omega_{b})/T_{c}}-\frac{\rho_{\beta\beta}^{ss}}{\rho_{bb}^{ss}}\right)=\nonumber \\
-\left(\omega_{a}-\omega_{\alpha}\right)\rho_{\alpha\alpha}^{ss}\Gamma-\left(\omega_{\beta}-\omega_{b}\right)\Gamma_{c}\left(1+\bar{N}_{c}\right)\rho_{bb}^{ss}\left(\frac{\gamma_{h}\bar{n}_{h}\Gamma\gamma_{c}\left(\bar{n}_{c}+1\right)}{\left(1+\bar{N}_{c}\right)\Gamma_{c}\left[\Gamma\left\{ \gamma_{c}\left(\bar{n}_{c}+1\right)+\gamma_{h}\left(\bar{n}_{h}+1\right)\right\} +\gamma_{h}\gamma_{c}\bar{n}_{c}\left(1+\bar{n}_{h}\right)\right]}\right)=\nonumber \\
-\frac{\rho_{bb}^{ss}\gamma_{h}\bar{n}_{h}\gamma_{c}\left(\bar{n}_{c}+1\right)\Gamma}{\Gamma\left[\gamma_{c}\left(\bar{n}_{c}+1\right)+\gamma_{h}\left(\bar{n}_{h}+1\right)\right]+\gamma_{h}\gamma_{c}\bar{n}_{c}\left(1+\bar{n}_{h}\right)}\left[\omega_{a}-\omega_{\alpha}+\omega_{\beta}-\omega_{b}\right],\\
-\frac{J_{c}}{J_{h}}=\frac{\omega_{a}-\omega_{\alpha}+\omega_{\beta}-\omega_{b}}{\omega_{a}-\omega_{b}}=1+\frac{\omega_{\beta}-\omega_{\alpha}}{\omega_{a}-\omega_{b}},
\end{gather}
where $\omega_{a}-\omega_{b}$ $\left(\omega_{\alpha}-\omega_{\beta}\right)$
is the energy of the absorbed (emitted) quanta from the hot bath (to
the RC/circuit). Therefore they are equivalent to $\omega_{abs}(\omega_{rc})$.
Using this paper notation, 

\begin{gather*}
J_{h}\rightarrow J_{abs},\\
J_{c}\rightarrow J_{loss},\\
\omega_{\beta}-\omega_{\alpha}\rightarrow-\omega_{rc},\\
\omega_{a}-\omega_{b}\rightarrow\omega_{abs},\\
T_{h}\rightarrow T_{abs},\\
T_{c}\rightarrow T_{loss},
\end{gather*}
we obtain Eq. 3 in the main text. A similar analysis can be done for
the coherence-assisted biological quantum heat engine model proposed
also in the same paper and to the model proposed on \cite{scully2011quantum}.

\begin{figure}
\label{fig:pnas model}

\begin{centering}
\includegraphics[scale=0.5]{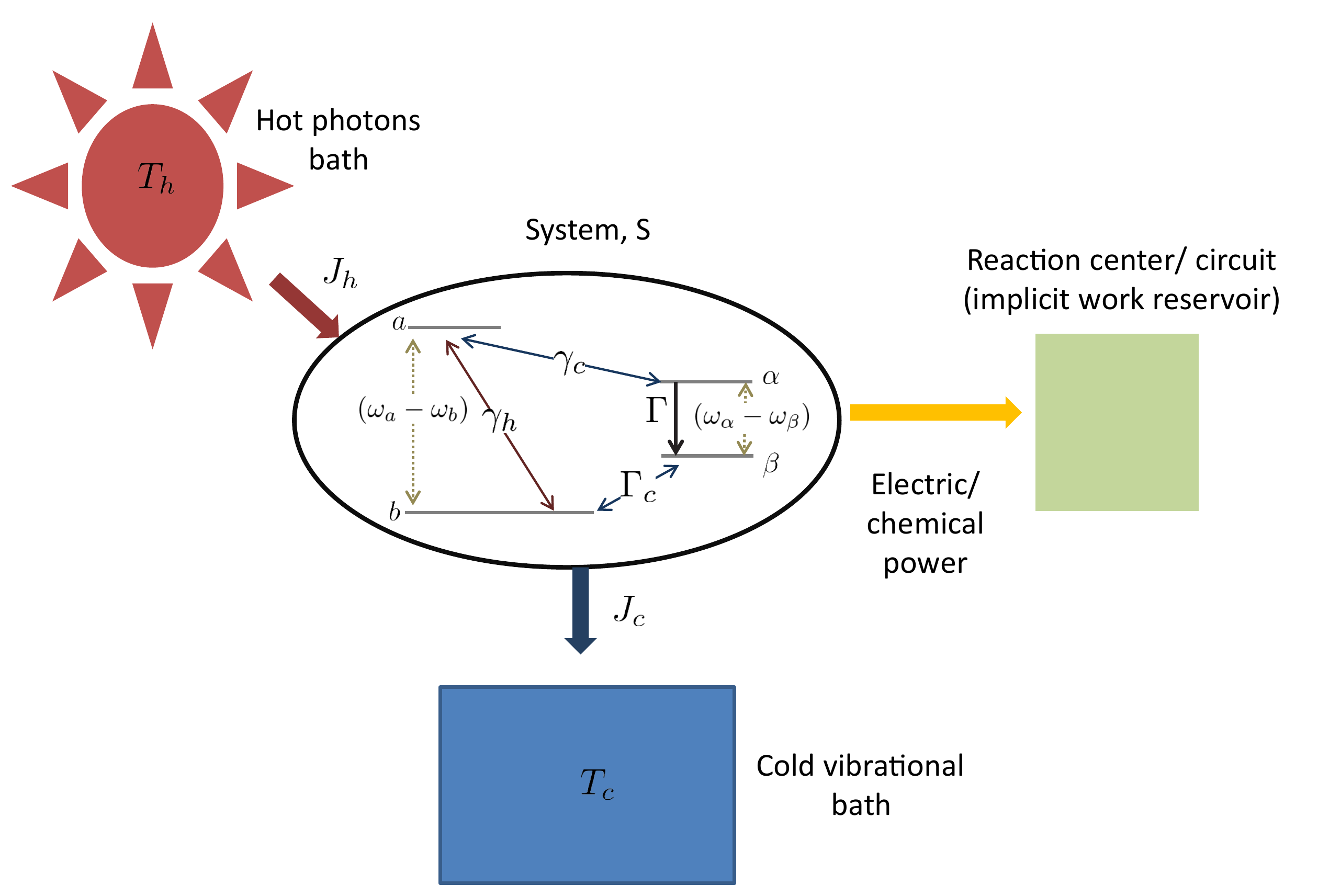}
\par\end{centering}

\caption{Biological quantum heat engine model from \cite{dorfman2013photosynthetic}. }
\end{figure}

2) We consider the photocell model proposed in \cite{creatore2013efficient}.
It consists of a five level system coupled to a hot bath, a cold bath
and to the reaction center/circuit (also termed ``the load''). $T_{h(c)}$
is the hot (cold) bath temperature. The decay rates are shown in Figure S2. For the sake of simplicity we assume there is no acceptor-to-donor
recombination ($\chi=0$, in the original paper notation). The equations
of motion are

\begin{gather}
\dot{\rho}_{\alpha\alpha}=\gamma_{c}\left[\left(1+n_{2c}\right)\rho_{x2x2}-n_{2c}\rho_{\alpha\alpha}\right]-\Gamma\rho_{\alpha\alpha},\nonumber \\
\dot{\rho}_{x2x2}=\gamma_{x}\left[(1+n_{x})\rho_{x1x1}-n_{x}\rho_{x2x2}\right]-\gamma_{c}\left[(1+n_{2c})\rho_{x2x2}-n_{2c}\rho_{\alpha\alpha}\right],\nonumber \\
\dot{\rho}_{bb}=-\left[\gamma_{h}n_{h}+\Gamma_{c}N_{c}\right]\rho_{bb}+\gamma_{h}\left(n_{h}+1\right)\rho_{x1x1}+\Gamma_{c}(N_{c}+1)\rho_{\beta\beta},\nonumber \\
\dot{\rho}_{x1x1}=-\gamma_{x}\left[(1+n_{x})\rho_{x1x1}-n_{x}\rho_{x2x2}\right]-\gamma_{h}\left[(1+n_{h})\rho_{x1x1}-n_{h}\rho_{bb}\right],\nonumber \\
\rho_{x1x1}+\rho_{x2x2}+\rho_{bb}+\rho_{\alpha\alpha}+\rho_{\beta\beta}=1.\label{eq:meq2}
\end{gather}
where we have kept the original paper notation. $\rho_{ii}$ is the
level population of state $i$ and $n_{i}$ or $N_{i}$ are the relevant
i- bath mode population. For details on Eq. \ref{eq:meq2} derivation,
we refer the reader to the original paper. The steady state populations
are

\begin{gather}
\frac{\rho_{x2x2}^{ss}}{\rho_{\alpha\alpha}^{ss}}=\frac{\Gamma+\gamma_{c}n_{2c}}{\gamma_{c}\left(1+n_{2c}\right)},\\
\frac{\rho_{x1x1}^{ss}}{\rho_{x2x2}^{ss}}=\frac{\gamma_{x}n_{x}+\gamma c\left(1+n_{2c}\right)-\gamma_{c}n_{2c}\frac{\gamma_{c}\left(1+n_{2c}\right)}{\Gamma+\gamma_{c}n_{2c}}}{\gamma_{x}\left(1+n_{x}\right)}=\frac{\gamma_{x}n_{x}\left(\Gamma+\gamma_{c}n_{2c}\right)+\gamma c\left(1+n_{2c}\right)\Gamma}{\gamma_{x}\left(1+n_{x}\right)\left(\Gamma+\gamma_{c}n_{2c}\right)},\\
\frac{\rho_{x1x1}^{ss}}{\rho_{bb}^{ss}}=\frac{\gamma_{h}n_{h}\left[\gamma_{x}n_{x}\left(\Gamma+\gamma_{c}n_{2c}\right)+\gamma_{c}(1+n_{2c})\Gamma\right]}{\gamma_{x}\gamma_{c}\Gamma\left(1+n_{x}\right)\left(1+n_{2c}\right)+\gamma_{h}\left(1+n_{h}\right)\left[\gamma_{x}n_{x}\left(\Gamma+\gamma_{c}n_{2c}\right)+\gamma_{c}(1+n_{2c})\Gamma\right]},\\
\frac{\rho_{\beta\beta}^{ss}}{\rho_{bb}^{ss}}=e^{-(\omega_{\beta}-\omega_{b})/T_{c}}+\frac{\gamma_{h}n_{h}\gamma_{x}\gamma_{c}\Gamma\left(1+n_{x}\right)\left(1+n_{2c}\right)}{\Gamma_{c}\left(1+N_{c}\right)\left\{ \gamma_{x}\gamma_{c}\Gamma\left(1+n_{x}\right)\left(1+n_{2c}\right)+\gamma_{h}\left(1+n_{h}\right)\left[\gamma_{x}n_{x}\left(\Gamma+\gamma_{c}n_{2c}\right)+\gamma_{c}\left(1+n_{2c}\right)\Gamma\right]\right\} }.
\end{gather}

Using Eq. \ref{eq:heatcurrents} the steady state heat currents are
obtained,

\begin{gather}
J_{h}=\left(\omega_{x1}-\omega_{b}\right)\left(1+\bar{n}_{h}\right)\gamma_{h}\rho_{bb}^{ss}\left(e^{-(\omega_{x1}-\omega_{b})/T_{h}}-\frac{\rho_{x1x1}^{ss}}{\rho_{bb}^{ss}}\right)=\nonumber \\
\left(\omega_{x1}-\omega_{b}\right)\frac{\rho_{bb}^{ss}\gamma_{h}n_{h}\gamma_{x}\gamma_{c}\Gamma\left(1+n_{x}\right)\left(1+n_{2c}\right)}{\gamma_{x}\gamma_{c}\Gamma\left(1+n_{x}\right)\left(1+n_{2c}\right)+\gamma_{h}\left(1+n_{h}\right)\left[\gamma_{x}n_{x}\left(\Gamma+\gamma_{c}n_{2c}\right)+\gamma_{c}\left(1+n_{2c}\right)\Gamma\right]},
\end{gather}

\begin{gather}
J_{c}=\left(\omega_{x1}-\omega_{x2}\right)\left(1+\bar{n}_{x}\right)\gamma_{x}\rho_{x2x2}^{ss}\left(e^{-(\omega_{x1}-\omega_{x2})/T_{c}}-\frac{\rho_{x1x1}^{ss}}{\rho_{x2x2}^{ss}}\right)+\nonumber \\
\left(\omega_{x2}-\omega_{\alpha}\right)\left(1+\bar{n}_{2c}\right)\gamma_{c}\rho_{\alpha\alpha}^{ss}\left(e^{-(\omega_{x2}-\omega_{\alpha})/T_{c}}-\frac{\rho_{x2x2}^{ss}}{\rho_{\alpha\alpha}^{ss}}\right)+\left(\omega_{\beta}-\omega_{b}\right)\left(1+\bar{N}_{c}\right)\Gamma_{c}\rho_{bb}^{ss}\left(e^{-(\omega_{\beta}-\omega_{b})/T_{c}}-\frac{\rho_{\beta\beta}^{ss}}{\rho_{bb}^{ss}}\right)=\nonumber \\
-\frac{\rho_{bb}^{ss}\gamma_{h}n_{h}\gamma_{x}\gamma_{c}\Gamma\left(1+n_{x}\right)\left(1+n_{2c}\right)}{\gamma_{x}\gamma_{c}\Gamma\left(1+n_{x}\right)\left(1+n_{2c}\right)+\gamma_{h}\left(1+n_{h}\right)\left[\gamma_{x}n_{x}\left(\Gamma+\gamma_{c}n_{2c}\right)+\gamma_{c}\left(1+n_{2c}\right)\Gamma\right]}\left(\omega_{x1}+\omega_{\beta}-\omega_{\alpha}-\omega_{b}\right),
\end{gather}

\begin{equation}
\frac{-J_{c}}{J_{h}}=\frac{\omega_{x1}+\omega_{\beta}-\omega_{\alpha}-\omega_{b}}{\omega_{x1}-\omega_{b}}=1+\frac{\omega_{\beta}-\omega_{\alpha}}{\omega_{x1}-\omega_{b}},
\end{equation}
where $\omega_{x1}-\omega_{b}$ ($\omega_{\alpha}-\omega_{\beta}$)
is the energy of the absorbed (emitted) quanta from the hot bath (to
the RC/circuit), therefore equivalent to $\omega_{abs}(\omega_{rc})$.
Using this paper notation,

\begin{gather*}
J_{h}\rightarrow J_{abs},\\
J_{c}\rightarrow J_{loss},\\
\omega_{\beta}-\omega_{\alpha}\rightarrow-\omega_{rc},\\
\omega_{x1}-\omega_{b}\rightarrow\omega_{abs},\\
T_{h}\rightarrow T_{abs},\\
T_{c}\rightarrow T_{loss},
\end{gather*}
 we obtain Eq. 3 in the main text. 

\begin{figure}
\label{fig:prl model}

\begin{centering}
\includegraphics[scale=0.5]{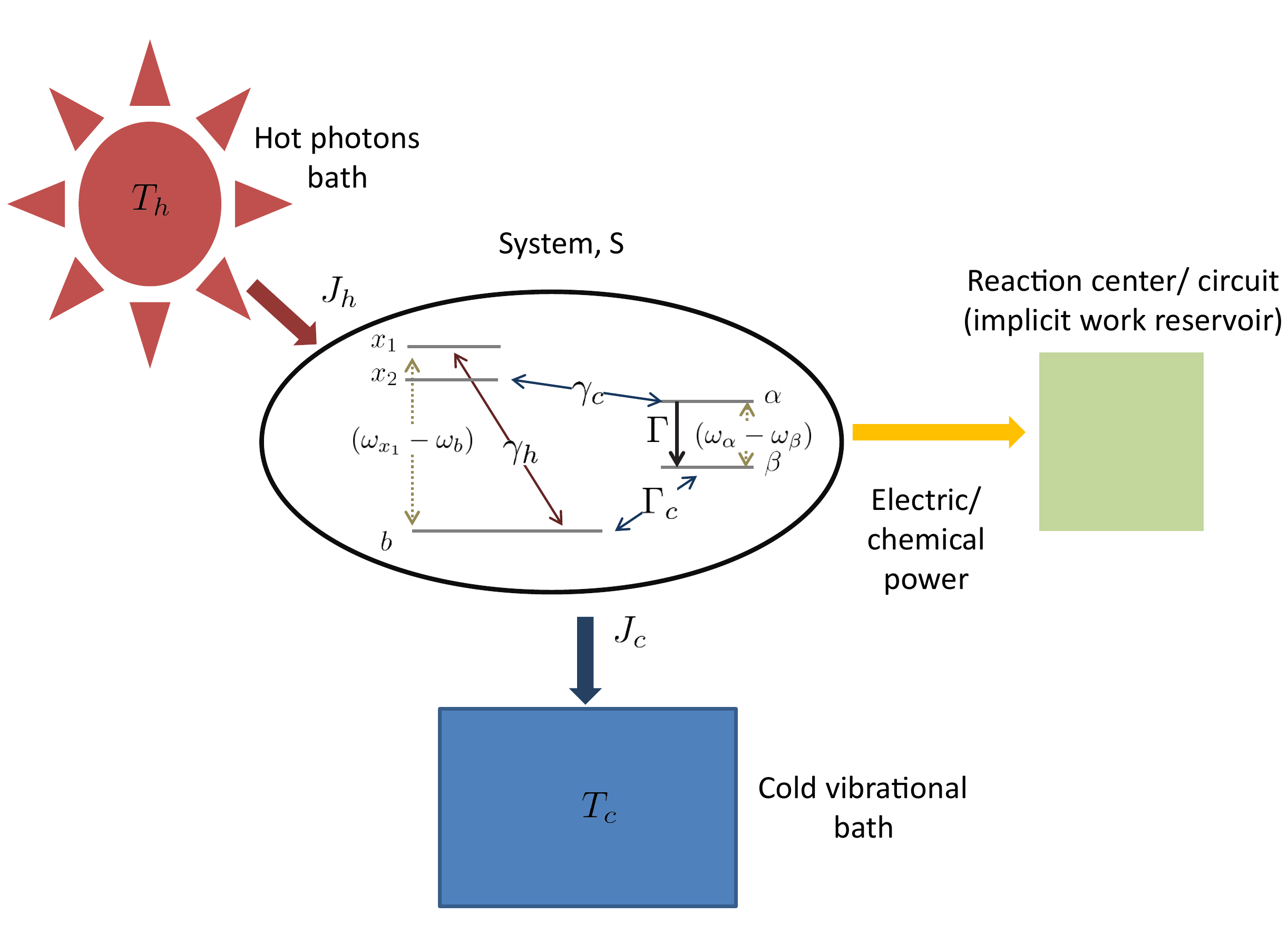}
\par\end{centering}

\caption{Photocell model proposed in \cite{creatore2013efficient}.}
\end{figure}

\subsection{FMO models}

We start by considering the model proposed on \cite{adolphs2006proteins}
for the Fenna-Mathews-Olson complex of a Prosthecochloris aestuarii.
Its dynamics is governed by the following Hamiltonian,

\begin{equation}
H_{FMO}+H_{FMO-vib}+H_{vib},
\end{equation}
where $H_{vib}$ is the free Hamiltonian for the vibrational degrees
of freedom of the pigments and proteins, which we assume to be at
equilibrium at a temperature $T_{loss}=300K.$ $H_{FMO}$ is the exciton
Hamiltonian, 

\begin{equation}
H_{FMO}=\sum_{m\in FMO}E_{m}|m\rangle\langle m|+\sum_{m\neq n\in FMO}V_{mn}|m\rangle\langle n|,
\end{equation}
where $|m\rangle$ is the excited state of the $m$ site, and the
sum is over all the FMO sites. $H_{FMO-vib}$ represents the interaction
between the excitons and the vibrations,

\begin{equation}
H_{FMO-vib}=\sum_{m\in FMO,\xi}k_{\xi}^{m}|m\rangle\langle m|\otimes Q_{\xi},
\end{equation}
where $Q_{\xi}$ operates   on the vibration degrees of freedom. All
the parameters for this Hamiltonian can be found on \cite{adolphs2006proteins}.

In order to thermodynamically analyze the FMO we complement the above
model with the following elements:

1) Energy transmission to the reaction center;

2) Absorption of thermal radiation by the antenna and its transmission
to the reaction center (RC), as well as the possibility for the FMO
sites to interact with the thermal radiation.

\subsubsection{Transmission of energy to the reaction center}

The transmission of energy to the reaction center is typically modeled
\cite{mohseni2008environment,rebentrost2009environment,plenio2008dephasing,killoran2014enhancing,alharbi2015theoretical,cao2009optimization,caruso2009highly}
as an irreversible decay term from the FMO site 3 to 8,

\begin{equation}
H_{Dec}=\sqrt{\Gamma_{3,8}}|8\rangle\langle3|.
\end{equation}

We use a typical value for this rate, $\Gamma_{3,8}=62.8/1.88\,cm^{-1}$\cite{mohseni2008environment,rebentrost2009environment,plenio2008dephasing,caruso2009highly}.

\subsubsection{Antenna and thermal radiation}

The antenna is composed of around 10,000 absorbing pigments \cite{valleau2014electromagnetic}.
As a simple model we consider the collective effect of these pigments
as an effective monochromatic antenna of frequency $\omega_{ant}=13333cm^{-1}$,
with an effective molecular transition dipole moment $\mu_{ant}=\sqrt{N}\mu_{ant,ind},$
where $N$ is the number of absorbing pigments and $\mu_{ant,ind}\sim5$
Debye, a typical value for a molecular transition dipole moment. 

Light absorption is governed by the antenna-radiation coupling Hamiltonian,

\begin{equation}
H_{ant-rad}=\mu_{ant}|ant\rangle\langle0|\otimes B_{abs}+h.c.,
\end{equation}
where $B_{abs}$ is an operator on the thermal radiation bath, $|ant\rangle$
is the antenna excited state and $|0\rangle$ is the ground state.
The FMO sites may also interact with the thermal radiation through
the Hamiltonian,

\begin{equation}
H_{FMO-rad}=\sum_{m}\mu_{FMO}|m\rangle\langle0|\otimes B_{abs}+h.c.,
\end{equation}
where $\mu_{FMO}=5.44$ Debye \cite{adolphs2006proteins}.

The transmission of the excitation from the antenna to the FMO is
assisted by the vibration degrees of freedom described by the Hamiltonian,

\begin{equation}
H_{ant-FMO}=\sum_{\xi,m\in FMO}\sqrt{\Gamma_{ant-FMO}}|m\rangle\langle ant|\otimes Q_{\xi}+h.c
\end{equation}
and we assume that $\Gamma_{ant-FMO}=\Gamma_{3,8}/10$.

Even though at the sun surface the thermal radiation emitted by the
sun is at equilibrium at the sun temperature, due to geometric considerations,
only a small fraction of those photons reaches the Earth. This is
quantified by a geometric factor $\lambda=2*10^{-5}$ equal to the
angle subtended by the Sun seen from the Earth. If $n_{T_{S}}\left[\omega\right]=(e^{\omega/T_{S}}-1)^{-1}$
photons of frequency $\omega,$ are emitted from the sun at temperature
$T_{s}$, only $\lambda n_{T_{s}}$ reach the Earth. This radiation
is no longer a thermal bath at the sun temperature, but rather is
a non-equilibrium bath at an effective temperature \cite{alicki2015non,landsberg1980thermodynamic},

\begin{equation}
e^{-\omega_{ant}/T_{abs}}=\frac{\lambda n_{T}\left[\omega_{ant}\right]}{\lambda n_{T}\left[\omega_{ant}\right]+1}\rightarrow T_{abs}\sim1356K.
\end{equation}

The dilution of the photon numbers turns the effective temperature,
$T_{abs}$, frequency dependent. Nevertheless, the frequency variation
between the antenna and the FMO site is small, therefore we assume
the same $T_{abs}$ for the antenna and the FMO sites.

\subsubsection*{Dynamic equations}

Collecting everything together, we can write the total Hamiltonian, 

\begin{equation}
H_{Tot}=H_{FMO}+H_{FMO-vib}+H_{vib}+H_{ant}+H_{ant-FMO}+H_{ant-rad}+H_{FMO-rad}+H_{rad}+H_{Dec},
\end{equation}
where $H_{ant(rad)}$ is the antenna (radiation) free Hamiltonian.

Using the standard Born-Markov approximation, the Lindblad equation
\cite{lindblad1976generators} for the FMO is numerically found, enabling
the calculation of the heat currents defined by Eq. \ref{eq:heatcurrents}.
$J_{abs}$ ($J_{loss}$) corresponds to the heat current between the
radiation (vibration) bath and the FMO.

\section{Simple models for the RC/circuit}

We consider a three level system (3LS), S, coupled to the reaction
center (RC) or electric circuit. The later is a reservoir of independent
quinones/sites, each of them represented by a single two level system
(TLS). Its ground state represents an empty quinone/site and the excited
state corresponds to a full quinone/site. Besides, the 3LS is coupled
to a photon (hot) bath and a vibrational (cold) bath (see Figure 3
in the main text). The total Hamiltonian is

\begin{equation}
H_{S}+H_{B}+H_{SB},
\end{equation}
where $H_{B}=H_{photons}+H_{phonons}$ is the baths free Hamiltonian.
The S-baths coupling Hamiltonian is given by

\begin{equation}
H_{SB}=S\otimes\left(B_{h}+B_{c}\right)=\sum_{\lambda}g_{h,\lambda}\left(|2\rangle\langle0|a_{\lambda}+|0\rangle\langle2|a_{\lambda}^{\dagger}\right)+\sum_{\lambda}g_{c,\lambda}\left(|2\rangle\langle1|b_{\lambda}+|1\rangle\langle2|b_{\lambda}^{\dagger}\right),\label{eq:hsb}
\end{equation}
where $a_{\lambda},$$a_{\lambda}^{\dagger}$ ($b_{\lambda},$ $b{}_{\lambda}^{\dagger}$)
are the annihilation and creation operator of photons (phonons) modes.
The S + RC/circuit Hamiltonian is 

\begin{gather}
H_{S}=H_{0}+H_{trans},\\
H_{0}=\omega_{abs}|2\rangle\langle2|+\frac{\omega_{rc}}{2}\left(|1\rangle\langle1|-|0\rangle\langle0|\right),
\end{gather}
where $H_{0}$ is the 3LS free Hamiltonian and $H_{trans}$ describes
the energy transfer to the RC/circuit. We compare between two possible
schemes: i) A decay transfer described by a non-hermitian $H_{trans}$;
ii) a Hamiltonian transfer, represented by a hermitic $H_{trans}$.

\subsection{Decay transfer}

The decay transfer is described by the following non-hermitian term,

\begin{equation}
H_{tranf}^{Dec}=\sqrt{\Gamma}|0\rangle\langle1|.
\end{equation}

As a first step we transform the S-bath interaction and the transfer
Hamiltonian to the interaction picture

\begin{gather}
H_{SB}\rightarrow e^{iH_{0}t}H_{SB}e^{-iH_{0}t},\quad H_{tranf}^{Dec}\rightarrow e^{iH_{0}t}H_{tranf}^{Dec}e^{-iH_{0}t}.\label{eq:dectransint}
\end{gather}
$H_{tranf}^{Dec}$ is a fictitious Hamiltonian due to its lack of
hermiticity, therefore can not form part of the rotation, $e^{iH_{0}t}$
, which has to be unitary. Besides, we derive the reduced dynamics
only for S. The operators in the interaction picture are: 

\begin{gather}
|2\rangle\langle0|\left[t\right]=e^{it\left(\omega_{abs}+\frac{\omega_{rc}}{2}\right)}|2\rangle\langle0|,\\
|2\rangle\langle1|\left[t\right]=e^{it\left(\omega_{abs}-\frac{\omega_{rc}}{2}\right)}|2\rangle\langle1|,\\
|0\rangle\langle1|\left[t\right]=e^{-it\omega_{rec}}|0\rangle\langle1|.
\end{gather}

Using the standard Born-Markov approximation, the Lindblad equation
\cite{lindblad1976generators} for S is obtained

\begin{gather}
\dot{\rho}_{22}=\nonumber \\
-\left\{ \Gamma_{h}\left(1+n_{h}\left[\omega_{abs}+\frac{\omega_{rc}}{2}\right]\right)+\Gamma_{c}\left(1+n_{c}\left[\omega_{abs}-\frac{\omega_{rc}}{2}\right]\right)\right\} \rho_{22}+\Gamma_{h}n_{h}\left[\omega_{abs}+\frac{\omega_{rc}}{2}\right]\rho_{00}+\Gamma_{c}n_{c}\left[\omega_{abs}-\frac{\omega_{rc}}{2}\right]\rho_{11},\\
\dot{\rho}_{11}=-\left\{ \Gamma+\Gamma_{c}n_{c}\left[\omega_{abs}-\frac{\omega_{rc}}{2}\right]\right\} \rho_{11}+\Gamma_{c}\left(1+n_{c}\left[\omega_{abs}-\frac{\omega_{rc}}{2}\right]\right)\rho_{22},\\
\dot{\rho}_{00}=-\Gamma_{h}n_{h}\left[\omega_{abs}+\frac{\omega_{rc}}{2}\right]\rho_{00}+\Gamma_{h}\left(1+n_{h}\left[\omega_{abs}+\frac{\omega_{rc}}{2}\right]\right)\rho_{22}+\Gamma\rho_{11},
\end{gather}

where $\Gamma_{i}$ and $n_{i}\left[\omega\right]$ are the decay
rate and $\omega$-mode population of the i-bath. The steady state
is 

\begin{gather}
\frac{\rho_{22}^{ss}}{\rho_{11}^{ss}}=\frac{\Gamma+\Gamma_{c}n_{c}\left[\omega_{abs}-\frac{\omega_{rc}}{2}\right]}{\Gamma_{c}\left(1+n_{c}\left[\omega_{abs}-\frac{\omega_{rc}}{2}\right]\right)},\\
\frac{\rho_{22}^{ss}}{\rho_{00}^{ss}}=e^{-\left(\omega_{abs}+\frac{\omega_{rc}}{2}\right)/T_{abs}}-\frac{\Gamma\rho_{11}^{ss}}{\Gamma_{h}\left(1+n_{h}\left[\omega_{abs}+\frac{\omega_{rc}}{2}\right]\right)\rho_{00}^{ss}},\\
\frac{\rho_{11}^{ss}}{\rho_{00}^{ss}}=\frac{\Gamma_{h}n_{h}\left[\omega_{abs}+\frac{\omega_{rc}}{2}\right]\Gamma_{c}\left(1+n_{c}\left[\omega_{abs}-\frac{\omega_{rc}}{2}\right]\right)}{\Gamma_{c}n_{c}\left[\omega_{abs}-\frac{\omega_{rc}}{2}\right]\Gamma_{h}\left(1+n_{h}\left[\omega_{abs}+\frac{\omega_{rc}}{2}\right]\right)+\Gamma\left\{ \Gamma_{h}\left(1+n_{h}\left[\omega_{abs}+\frac{\omega_{rc}}{2}\right]\right)+\Gamma_{c}\left(1+n_{c}\left[\omega_{abs}-\frac{\omega_{rc}}{2}\right]\right)\right\} },\\
\rho_{11}^{ss}=\frac{1}{1+\frac{\rho_{00}^{ss}}{\rho_{11}^{ss}}+\frac{\rho_{22}^{ss}}{\rho_{11}^{ss}}}=\frac{1}{1+2e^{\left(\omega_{abs}+\frac{\omega_{rc}}{2}\right)/T_{abs}}}.
\end{gather}

In the last equality, we assume for simplicity that all the zero
temperature decay rates are equal to the RC decay rate, $\Gamma_{c}=\Gamma_{h}=\Gamma$.
Using Eq. \ref{eq:heatcurrents} the heat currents at steady state
are obtained,
\begin{gather}
J_{abs}^{Dec}\equiv J_{h}^{Dec}=\left(\omega_{abs}+\frac{\omega_{rc}}{2}\right)\rho_{00}^{ss}\Gamma_{h}\left(1+n_{h}\left[\omega_{abs}+\frac{\omega_{rc}}{2}\right]\right)\left[e^{-\left(\omega_{abs}+\frac{\omega_{rc}}{2}\right)/T_{abs}}-\frac{\rho_{22}^{ss}}{\rho_{00}^{ss}}\right]=\left(\omega_{abs}+\frac{\omega_{rc}}{2}\right)\Gamma\rho_{11}^{ss},\\
J_{loss}^{Dec}\equiv J_{c}^{Dec}=\left(\omega_{abs}-\frac{\omega_{rc}}{2}\right)\rho_{11}^{ss}\Gamma_{c}\left(1+n_{c}\left[\omega_{abs}-\frac{\omega_{rc}}{2}\right]\right)\left[e^{-\left(\omega_{abs}-\frac{\omega_{rc}}{2}\right)/T_{loss}}-\frac{\rho_{22}^{ss}}{\rho_{11}^{ss}}\right]=-\left(\omega_{abs}-\frac{\omega_{rc}}{2}\right)\Gamma\rho_{11}^{ss}
\end{gather}

and by energy conservation the power is 

\begin{equation}
P^{Dec}=-J_{abs}^{Dec}-J_{loss}^{Dec}=-\omega_{rc}\Gamma\rho_{11}^{ss}<0.
\end{equation}
This model predicts that power is extracted independently of the baths
temperatures, in contradiction with the second law of thermodynamics
which forbids power extraction in the case of a single temperature,
$T_{loss}=T_{abs}$. As shown in Figure 4 of the main text, also for
$T_{loss}\neq T_{abs}$, $P^{Dec}$ diverges from the extracted power
predicted by a thermodynamically consistent model.

\subsection{Hamiltonian transfer}

Here we explicitly consider the RC/circuit and its coupling to S,
by considering $H_{trans}$ as an hermitic Hamiltonian. The RC/circuit
is composed of identical and independent two level systems (TLS).
The S + RC/circuit Hamiltonian is:

\begin{equation}
H_{S}=\omega_{abs}|2\rangle\langle2|+\frac{\omega_{rc}}{2}\left(|1\rangle\langle1|-|0\rangle\langle0|\right)+\sum_{k}^{j}\sqrt{\frac{\Gamma}{2j}}\left(\sigma_{-}^{k}|1\rangle\langle0|+\sigma_{+}^{k}|0\rangle\langle1|\right)+\omega_{rc}\sum_{k}\sigma_{z}^{k},
\end{equation}

where $j$ is the number of TLSs.

In order to find the energy that is being transferred to the RC/circuit,
we start by diagonalizing the S + RC circuit. This is achieved by
first applying the Holstein-Primakoff transformation \cite{holstein1940field},
that consist on the introduction of the following collective operators:

\begin{gather}
\sum_{k}\sigma_{-}^{k}=\left(\sqrt{2j-c^{\dagger}c}\right)c,\\
\sum_{k}\sigma_{+}^{k}=c^{\dagger}\left(\sqrt{2j-c^{\dagger}c}\right),\\
\sum_{k}\sigma_{z}^{k}=c^{\dagger}c-j.
\end{gather}

The new Hamiltonian is

\begin{equation}
H_{S}=\omega_{abs}|2\rangle\langle2|+\frac{\omega_{rc}}{2}\left(|1\rangle\langle1|-|0\rangle\langle0|\right)+\sqrt{\frac{\Gamma}{2j}}\left[\left(\sqrt{2j-c^{\dagger}c}\right)c|1\rangle\langle0|+c^{\dagger}\left(\sqrt{2j-c^{\dagger}c}\right)|0\rangle\langle1|\right]+\omega_{rc}\left(c^{\dagger}c-j\right).
\end{equation}

At this point, the modes are displaced, $c\rightarrow c-\sqrt{\epsilon}$,

\begin{gather}
H_{S}=\omega_{abs}|2\rangle\langle2|+\frac{\omega_{rc}}{2}\left(|1\rangle\langle1|-|0\rangle\langle0|\right)+\sqrt{\frac{\Gamma k\eta}{2j}}\left(c|1\rangle\langle0|+c^{\dagger}|0\rangle\langle1|\right)-\sqrt{\frac{\Gamma k\eta\epsilon}{2j}}\left(|1\rangle\langle0|+|0\rangle\langle1|\right)+\nonumber \\
\omega_{rc}\left(c^{\dagger}c-\sqrt{\epsilon}(c+c^{\dagger})+\epsilon-j\right),
\end{gather}
where $k=2j-\epsilon$ and $\eta=1-\frac{c^{\dagger}c-\sqrt{\epsilon}(c^{\dagger}+c)}{k}$.
We assume that the number of TLSs is large, $\frac{c^{\dagger}c-\sqrt{\epsilon}(c^{\dagger}+c)}{k}\ll1$.
The physical interpretation of this approximation is clarified below.
Under this assumptions, we expand $\sqrt{\eta}\approx1-\frac{c^{\dagger}c-\sqrt{\epsilon}(c^{\dagger}+c)}{2k}-\frac{\epsilon(c^{\dagger}+c)^{2}}{8k^{2}}$
and keep terms up to order $\frac{1}{\sqrt{j}},$

\begin{gather}
H_{S}=\omega_{abs}|2\rangle\langle2|+\frac{\omega_{rc}}{2}\left(|1\rangle\langle1|-|0\rangle\langle0|\right)-\sqrt{\frac{\Gamma k\epsilon}{2j}}\left(|1\rangle\langle0|+|0\rangle\langle1|\right)+\sqrt{\frac{\Gamma k}{2j}}\left(c|1\rangle\langle0|+c^{\dagger}|0\rangle\langle1|\right)-\nonumber \\
\frac{\epsilon}{2}\sqrt{\frac{\Gamma}{2jk}}\left(c^{\dagger}+c\right)\left(|1\rangle\langle0|+|0\rangle\langle1|\right)+\omega_{rc}\left(c^{\dagger}c-\sqrt{\epsilon}\left(c+c^{\dagger}\right)+\epsilon-j\right).
\end{gather}
 Setting $\epsilon=0$, the Hamiltonian is simplified to

\begin{gather}
H_{S}=\omega_{abs}|2\rangle\langle2|+\frac{\omega_{rc}}{2}\left(|1\rangle\langle1|-|0\rangle\langle0|\right)+\sqrt{\Gamma}\left(c|1\rangle\langle0|+c^{\dagger}|0\rangle\langle1|\right)+\omega_{rc}\left(c^{\dagger}c-j\right)\label{eq:hph}
\end{gather}
and the approximation to $\frac{c^{\dagger}c}{2j}\ll1$. Therefore,
we are just assuming that the total number of excitations in the RC/circuit
is very small compared to the number of quinones/sites, so energy
may always be transferred to the RC/circuit. From Eq. \ref{eq:hph},
we derive Eq. 9 in the main text,

\begin{equation}
H_{trasns}^{emi}=\sqrt{\Gamma}\left(c|1\rangle\langle0|+c^{\dagger}|0\rangle\langle1|\right)+\omega_{rc}\left(c^{\dagger}c-j\right).
\end{equation}

Next we diagonalize Eq. \ref{eq:hph}. The Hamiltonian eigenvectors
are

\begin{gather}
|+,n\rangle=\frac{1}{\sqrt{2}}\left(|1,n\rangle+|0,n+1\rangle\right),\\
|-,n\rangle=\frac{1}{\sqrt{2}}\left(|0,n+1\rangle-|1,n\rangle\right),\\
E_{\pm}=\omega_{rc}\left(n+\frac{1}{2}\right)\pm\frac{\Omega_{n}}{2}-j\omega_{rc},\\
H_{S}=\omega_{abs}|2\rangle\langle2|+\omega_{rc}\left(\tilde{c}^{\dagger}\tilde{c}+\frac{1}{2}\sum_{n}\left(\left|+,n\right\rangle \langle+,n|+|-,n\rangle\langle-,n|\right)-j\right)+\frac{\Omega_{n}}{2}\tilde{\sigma}_{z},
\end{gather}
where $\tilde{c}^{\dagger}(\tilde{c})$ is the creation (annihilation)
operator in the new basis and $\Omega_{n}=2\sqrt{\Gamma(n+1)}$. The
inverse transformations are

\begin{gather}
|1,n\rangle=\frac{1}{\sqrt{2}}\left(|+,n\rangle-|-,n\rangle\right),\\
|0,n+1\rangle=\frac{1}{\sqrt{2}}\left(|+,n\rangle+|-,n\rangle\right).
\end{gather}

Rewriting the S-bath Hamiltonian, Eq. \ref{eq:hsb} , in the new basis,

\begin{gather}
|2\rangle\langle0|=\sum_{n}\frac{1}{\sqrt{2}}\left(|2,n+1\rangle\langle+,n|+|2,n+1\rangle\langle-,n|\right),\\
|2\rangle\langle1|=\sum_{n}\frac{1}{\sqrt{2}}\left(|2,n\rangle\langle+,n|-|2,n\rangle\langle-,n|\right),
\end{gather}
and transforming to the interaction picture,

\begin{gather}
H_{SB}\rightarrow e^{iH_{S}t}H_{SB}e^{-iH_{S}t},\\
|2\rangle\langle0|\left[t\right]=\sum_{n}\frac{1}{\sqrt{2}}\left(e^{it\left(\omega_{abs}+\frac{\omega_{rc}}{2}-\frac{\Omega_{n}}{2}\right)}|2,n+1\rangle\langle+,n|+e^{it\left(\omega_{abs}+\frac{\omega_{rc}}{2}+\frac{\Omega_{n}}{2}\right)}|2,n+1\rangle\langle-,n|\right),\\
|2\rangle\langle1|\left[t\right]=\sum_{n}\frac{1}{\sqrt{2}}\left(e^{it\left(\omega_{abs}-\frac{\omega_{rc}}{2}-\frac{\Omega_{n}}{2}\right)}|2,n\rangle\langle+,n|-e^{it\left(\omega_{abs}-\frac{\omega_{rc}}{2}+\frac{\Omega_{n}}{2}\right)}|2,n\rangle\langle-,n|\right).
\end{gather}

In contrast to the decay transfer scheme (Eq. \ref{eq:dectransint}),
here $H_{trans}^{Ham}$ is hermitian and we derive the reduced dynamics
for the S + RC/circuit. Therefore $H_{trans}^{Ham}$ is included in
the rotation, $e^{iH_{S}t}$.

Using the standard Born-Markov approximation, the Lindblad equation
\cite{lindblad1976generators} for S + RC/circuit is obtained, and
from it the evolution equations are derived, 
\begin{gather}
\dot{\rho}_{+,n}=\frac{1}{2}\left\{ -\Gamma_{h}n_{h}\left[\omega_{abs}+\frac{\omega_{rc}-\Omega_{n}}{2}\right]\rho_{+,n}+\Gamma_{h}\left(1+n_{h}\left[\omega_{abs}+\frac{\omega_{rc}-\Omega_{n}}{2}\right]\right)\rho_{2,n+1}\right\} +\nonumber \\
\frac{1}{2}\left\{ -\Gamma_{c}n_{c}\left[\omega_{abs}-\frac{\omega_{rc}+\Omega_{n}}{2}\right]\rho_{+,n}+\Gamma_{c}\left(1+n_{c}\left[\omega_{abs}-\frac{\omega_{rc}+\Omega_{n}}{2}\right]\right)\rho_{2,n}\right\} ,\nonumber \\
\dot{\rho}_{-,n}=\frac{1}{2}\left\{ -\Gamma_{h}n_{h}\left[\omega_{abs}+\frac{\omega_{rc}+\Omega_{n}}{2}\right]\rho_{-,n}+\Gamma_{h}\left(1+n_{h}\left[\omega_{abs}+\frac{\omega_{rc}+\Omega_{n}}{2}\right]\right)\rho_{2,n+1}\right\} +\nonumber \\
\frac{1}{2}\left\{ -\Gamma_{c}n_{c}\left[\omega_{abs}-\frac{\omega_{rc}-\Omega_{n}}{2}\right]\rho_{-,n}+\Gamma_{c}\left(1+n_{c}\left[\omega_{abs}-\frac{\omega_{rc}-\Omega_{n}}{2}\right]\right)\rho_{2,n}\right\} ,\nonumber \\
\dot{\rho}_{2,n}=\frac{1}{2}\left\{ \Gamma_{h}n_{h}\left[\omega_{abs}+\frac{\omega_{rc}-\Omega_{n-1}}{2}\right]\rho_{+,n-1}-\Gamma_{h}\left(1+n_{h}\left[\omega_{abs}+\frac{\omega_{rc}-\Omega_{n-1}}{2}\right]\right)\rho_{2,n}\right\} +\nonumber \\
\frac{1}{2}\left\{ \Gamma_{h}n_{h}\left[\omega_{abs}+\frac{\omega_{rc}+\Omega_{n-1}}{2}\right]\rho_{-,n-1}-\Gamma_{h}\left(1+n_{h}\left[\omega_{abs}+\frac{\omega_{rc}+\Omega_{n-1}}{2}\right]\right)\rho_{2,n}\right\} +\nonumber \\
\frac{1}{2}\left\{ \Gamma_{c}n_{c}\left[\omega_{abs}-\frac{\omega_{rc}+\Omega_{n}}{2}\right]\rho_{+,n}-\Gamma_{c}\left(1+n_{c}\left[\omega_{abs}-\frac{\omega_{rc}+\Omega_{n}}{2}\right]\right)\rho_{2,n}\right\} +\label{eq:eqom}\\
\frac{1}{2}\left\{ \Gamma_{c}n_{c}\left[\omega_{abs}-\frac{\omega_{rc}-\Omega_{n}}{2}\right]\rho_{-,n}-\Gamma_{c}\left(1+n_{c}\left[\omega_{abs}-\frac{\omega_{rc}-\Omega_{n}}{2}\right]\right)\rho_{2,n}\right\} ,\nonumber 
\end{gather}
where $\rho_{i}$ is the population of the combined state $i$ (S
+ RC/circuit), $\Gamma_{i}$ and $n_{i}\left[\omega\right]$ are the
decay rate and $\omega$-mode population of the i-bath. The equations
for the off-diagonal terms are decoupled from the populations and
we assume them to be zero. If the coupling between the 3LS and the
RC/circuit is weak, $\omega_{rc}\gg\Omega_{n}$, it can be assumed
that they are in a product state. Moreover, if the coupling spectrum
is approximately flat in frequency windows of size $\Omega_{n}-\Omega_{n-1}$,
the 3LS steady state is

\begin{gather}
\frac{\rho_{2}^{ss}}{\rho_{+}^{ss}}=\frac{\rho_{2}^{ss}}{\rho_{-}^{ss}}=\frac{\Gamma_{h}n_{h}\left[\omega_{abs}+\frac{\omega_{rc}}{2}\right]+\Gamma_{c}n_{c}\left[\omega_{abs}-\frac{\omega_{rc}}{2}\right]}{\Gamma_{h}\left(1+n_{h}\left[\omega_{abs}+\frac{\omega_{rc}}{2}\right]\right)+\Gamma_{c}\left(1+n_{c}\left[\omega_{abs}-\frac{\omega_{rc}}{2}\right]\right)}=\frac{n_{h}\left[\omega_{abs}+\frac{\omega_{rc}}{2}\right]+n_{c}\left[\omega_{abs}-\frac{\omega_{rc}}{2}\right]}{2+n_{h}\left[\omega_{abs}+\frac{\omega_{rc}}{2}\right]+n_{c}\left[\omega_{abs}-\frac{\omega_{rc}}{2}\right]},
\end{gather}
\\
\begin{gather}
\rho_{+}^{ss}=\frac{2+n_{h}\left[\omega_{abs}+\frac{\omega_{rc}}{2}\right]+n_{c}\left[\omega_{abs}-\frac{\omega_{rc}}{2}\right]}{4+3n_{h}\left[\omega_{abs}+\frac{\omega_{rc}}{2}\right]+3n_{c}\left[\omega_{abs}-\frac{\omega_{rc}}{2}\right]}.
\end{gather}

For the sake of simplicity we have assumed in the last equality that
the zero temperature decay rates of both baths are the same as the
RC/circuit coupling strength, $\Gamma_{h}=\Gamma_{c}=\Gamma$. From
Eqs. \ref{eq:eqom} an evolution equation for the RC/circuit can be
written,

\begin{equation}
\dot{\rho}_{n}=\dot{\rho}_{+,n}+\dot{\rho}_{-,n}+\dot{\rho}_{2,n}=r\rho_{n+1}+s\rho_{n-1}-(r+s)\rho_{n},
\end{equation}
which is a ``birth-death process'' \cite{van1992stochastic}, where
$s(r)$ is the birth (death) rate,

\begin{equation}
s=\frac{1}{2}\Gamma_{h}n_{h}\left[\omega_{abs}+\frac{\omega_{rc}}{2}\right]\left(\rho_{+}+\rho_{-}\right),\quad r=\Gamma_{h}\left(1+n_{h}\left[\omega_{abs}+\frac{\omega_{rc}}{2}\right]\right)\rho_{2}.
\end{equation}

The energy change of the RC/circuit evolves as

\begin{equation}
\omega_{rc}\langle\dot{n}\rangle=\left(s-r\right)\omega_{rc},
\end{equation}

which is equal to $-P^{Ham}$ (the used sign convention can be found
below Eq. 1 in the main text).

Thus, $s>r$ is required in order to increase the RC/circuit energy.
At the 3LS steady state, this implies,

\begin{gather}
s-r=\frac{\Gamma\left(1+n_{c}\left[\omega_{abs}-\frac{\omega_{rc}}{2}\right]\right)\left(1+n_{h}\left[\omega_{abs}+\frac{\omega_{rc}}{2}\right]\right)}{4+3n_{h}\left[\omega_{abs}+\frac{\omega_{rc}}{2}\right]+3n_{c}\left[\omega_{abs}-\frac{\omega_{rc}}{2}\right]}\left(e^{-\left(\omega_{abs}+\frac{\omega_{rc}}{2}\right)/T_{abs}}-e^{-\left(\omega_{abs}-\frac{\omega_{rc}}{2}\right)/T_{loss}}\right)=\nonumber \\
K_{1}\left(e^{-\left(\omega_{abs}+\frac{\omega_{rc}}{2}\right)/T_{abs}}-e^{\left(\omega_{abs}-\frac{\omega_{rc}}{2}\right)/T_{loss}}\right)>0,
\end{gather}
where $K_{1}=\frac{\Gamma\left(1+n_{c}\left[\omega_{abs}-\frac{\omega_{rc}}{2}\right]\right)\left(1+n_{h}\left[\omega_{abs}+\frac{\omega_{rc}}{2}\right]\right)}{4+3n_{h}\left[\omega_{abs}+\frac{\omega_{rc}}{2}\right]+3n_{c}\left[\omega_{abs}-\frac{\omega_{rc}}{2}\right]}>0$
and the energy gain condition is 

\begin{equation}
\frac{T_{loss}}{T_{abs}}<\frac{\omega_{abs}-\frac{\omega_{rc}}{2}}{\omega_{abs}+\frac{\omega_{rc}}{2}}=\frac{2\omega_{abs}-\omega_{rc}}{2\omega_{abs}+\omega_{rc}}.
\end{equation}

Using Eq. \ref{eq:heatcurrents} the heat currents at steady state
are obtained,

\begin{gather}
J_{abs}^{Ham}\equiv J_{h}^{Ham}=\left(\omega_{abs}+\frac{\omega_{rc}}{2}\right)(s-r),\\
J_{loss}^{Ham}\equiv J_{c}^{Ham}=-\left(\omega_{abs}-\frac{\omega_{rc}}{2}\right)(s-r).
\end{gather}

%

\begin{thebibliography}{41}
\expandafter\ifx\csname natexlab\endcsname\relax\def\natexlab#1{#1}\fi
\expandafter\ifx\csname bibnamefont\endcsname\relax
  \def\bibnamefont#1{#1}\fi
\expandafter\ifx\csname bibfnamefont\endcsname\relax
  \def\bibfnamefont#1{#1}\fi
\expandafter\ifx\csname citenamefont\endcsname\relax
  \def\citenamefont#1{#1}\fi
\expandafter\ifx\csname url\endcsname\relax
  \def\url#1{\texttt{#1}}\fi
\expandafter\ifx\csname urlprefix\endcsname\relax\def\urlprefix{URL }\fi
\providecommand{\bibinfo}[2]{#2}
\providecommand{\eprint}[2][]{\url{#2}}

\bibitem[{\citenamefont{Blankenship}(2013)}]{blankenship2013molecular}
\bibinfo{author}{\bibfnamefont{R.~E.} \bibnamefont{Blankenship}},
  \emph{\bibinfo{title}{Molecular mechanisms of photosynthesis}}
  (\bibinfo{publisher}{John Wiley \& Sons}, \bibinfo{year}{2013}).

\bibitem[{\citenamefont{Nelson}(2003)}]{nelson2003physics}
\bibinfo{author}{\bibfnamefont{J.}~\bibnamefont{Nelson}},
  \emph{\bibinfo{title}{The physics of solar cells}}, vol.~\bibinfo{volume}{1}
  (\bibinfo{publisher}{World Scientific}, \bibinfo{year}{2003}).

\bibitem[{\citenamefont{W{\"u}rfel and W{\"u}rfel}(2009)}]{wurfel2009physics}
\bibinfo{author}{\bibfnamefont{P.}~\bibnamefont{W{\"u}rfel}} \bibnamefont{and}
  \bibinfo{author}{\bibfnamefont{U.}~\bibnamefont{W{\"u}rfel}},
  \emph{\bibinfo{title}{Physics of solar cells: from basic principles to
  advanced concepts}} (\bibinfo{publisher}{John Wiley \& Sons},
  \bibinfo{year}{2009}).

\bibitem[{\citenamefont{Blankenship et~al.}(2011)\citenamefont{Blankenship,
  Tiede, Barber, Brudvig, Fleming, Ghirardi, Gunner, Junge, Kramer, Melis
  et~al.}}]{blankenship2011comparing}
\bibinfo{author}{\bibfnamefont{R.~E.} \bibnamefont{Blankenship}},
  \bibinfo{author}{\bibfnamefont{D.~M.} \bibnamefont{Tiede}},
  \bibinfo{author}{\bibfnamefont{J.}~\bibnamefont{Barber}},
  \bibinfo{author}{\bibfnamefont{G.~W.} \bibnamefont{Brudvig}},
  \bibinfo{author}{\bibfnamefont{G.}~\bibnamefont{Fleming}},
  \bibinfo{author}{\bibfnamefont{M.}~\bibnamefont{Ghirardi}},
  \bibinfo{author}{\bibfnamefont{M.}~\bibnamefont{Gunner}},
  \bibinfo{author}{\bibfnamefont{W.}~\bibnamefont{Junge}},
  \bibinfo{author}{\bibfnamefont{D.~M.} \bibnamefont{Kramer}},
  \bibinfo{author}{\bibfnamefont{A.}~\bibnamefont{Melis}},
  \bibnamefont{et~al.}, \bibinfo{journal}{science}
  \textbf{\bibinfo{volume}{332}}, \bibinfo{pages}{805} (\bibinfo{year}{2011}).

\bibitem[{\citenamefont{Alicki et~al.}(2016)\citenamefont{Alicki,
  Gelbwaser-Klimovsky, and Szczygielski}}]{alicki2015solar}
\bibinfo{author}{\bibfnamefont{R.}~\bibnamefont{Alicki}},
  \bibinfo{author}{\bibfnamefont{D.}~\bibnamefont{Gelbwaser-Klimovsky}},
  \bibnamefont{and}
  \bibinfo{author}{\bibfnamefont{K.}~\bibnamefont{Szczygielski}},
  \bibinfo{journal}{Journal of Physics A: Mathematical and Theoretical}
  \textbf{\bibinfo{volume}{49}}, \bibinfo{pages}{015002}
  (\bibinfo{year}{2016}),
  \urlprefix\url{http://stacks.iop.org/1751-8121/49/i=1/a=015002}.

\bibitem[{\citenamefont{Einax and Nitzan}(2014)}]{einax2014network}
\bibinfo{author}{\bibfnamefont{M.}~\bibnamefont{Einax}} \bibnamefont{and}
  \bibinfo{author}{\bibfnamefont{A.}~\bibnamefont{Nitzan}},
  \bibinfo{journal}{The Journal of Physical Chemistry C}
  \textbf{\bibinfo{volume}{118}}, \bibinfo{pages}{27226}
  (\bibinfo{year}{2014}).

\bibitem[{\citenamefont{Kondepudi and Prigogine}(2014)}]{kondepudi2014modern}
\bibinfo{author}{\bibfnamefont{D.}~\bibnamefont{Kondepudi}} \bibnamefont{and}
  \bibinfo{author}{\bibfnamefont{I.}~\bibnamefont{Prigogine}},
  \emph{\bibinfo{title}{Modern thermodynamics: from heat engines to dissipative
  structures}} (\bibinfo{publisher}{John Wiley \& Sons}, \bibinfo{year}{2014}).

\bibitem[{\citenamefont{Gelbwaser-Klimovsky
  et~al.}(2015)\citenamefont{Gelbwaser-Klimovsky, Niedenzu, and
  Kurizki}}]{gelbwaser2015thermodynamics}
\bibinfo{author}{\bibfnamefont{D.}~\bibnamefont{Gelbwaser-Klimovsky}},
  \bibinfo{author}{\bibfnamefont{W.}~\bibnamefont{Niedenzu}}, \bibnamefont{and}
  \bibinfo{author}{\bibfnamefont{G.}~\bibnamefont{Kurizki}},
  \bibinfo{journal}{Advances In Atomic, Molecular, and Optical Physics}
  \textbf{\bibinfo{volume}{64}}, \bibinfo{pages}{329} (\bibinfo{year}{2015}).

\bibitem[{\citenamefont{Shockley and Queisser}(1961)}]{shockley1961detailed}
\bibinfo{author}{\bibfnamefont{W.}~\bibnamefont{Shockley}} \bibnamefont{and}
  \bibinfo{author}{\bibfnamefont{H.~J.} \bibnamefont{Queisser}},
  \bibinfo{journal}{Journal of applied physics} \textbf{\bibinfo{volume}{32}},
  \bibinfo{pages}{510} (\bibinfo{year}{1961}).

\bibitem[{\citenamefont{Landsberg and
  Tonge}(1980)}]{landsberg1980thermodynamic}
\bibinfo{author}{\bibfnamefont{P.}~\bibnamefont{Landsberg}} \bibnamefont{and}
  \bibinfo{author}{\bibfnamefont{G.}~\bibnamefont{Tonge}},
  \bibinfo{journal}{Journal of Applied Physics} \textbf{\bibinfo{volume}{51}},
  \bibinfo{pages}{R1} (\bibinfo{year}{1980}).

\bibitem[{\citenamefont{Knox and Parson}(2007)}]{knox2007entropy}
\bibinfo{author}{\bibfnamefont{R.~S.} \bibnamefont{Knox}} \bibnamefont{and}
  \bibinfo{author}{\bibfnamefont{W.~W.} \bibnamefont{Parson}},
  \bibinfo{journal}{Biochimica et Biophysica Acta (BBA)-Bioenergetics}
  \textbf{\bibinfo{volume}{1767}}, \bibinfo{pages}{1189}
  (\bibinfo{year}{2007}).

\bibitem[{\citenamefont{Knox}(1969)}]{knox1969thermodynamics}
\bibinfo{author}{\bibfnamefont{R.~S.} \bibnamefont{Knox}},
  \bibinfo{journal}{Biophysical journal} \textbf{\bibinfo{volume}{9}},
  \bibinfo{pages}{1351} (\bibinfo{year}{1969}).

\bibitem[{\citenamefont{Parson}(1978)}]{parson1978thermodynamics}
\bibinfo{author}{\bibfnamefont{W.~W.} \bibnamefont{Parson}},
  \bibinfo{journal}{Photochemistry and photobiology}
  \textbf{\bibinfo{volume}{28}}, \bibinfo{pages}{389} (\bibinfo{year}{1978}).

\bibitem[{\citenamefont{Ross and Calvin}(1967)}]{ross1967thermodynamics}
\bibinfo{author}{\bibfnamefont{R.~T.} \bibnamefont{Ross}} \bibnamefont{and}
  \bibinfo{author}{\bibfnamefont{M.}~\bibnamefont{Calvin}},
  \bibinfo{journal}{Biophysical journal} \textbf{\bibinfo{volume}{7}},
  \bibinfo{pages}{595} (\bibinfo{year}{1967}).

\bibitem[{\citenamefont{Alicki and Gelbwaser-Klimovsky}(2015)}]{alicki2015non}
\bibinfo{author}{\bibfnamefont{R.}~\bibnamefont{Alicki}} \bibnamefont{and}
  \bibinfo{author}{\bibfnamefont{D.}~\bibnamefont{Gelbwaser-Klimovsky}},
  \bibinfo{journal}{New Journal of Physics} \textbf{\bibinfo{volume}{17}},
  \bibinfo{pages}{115012} (\bibinfo{year}{2015}).

\bibitem[{\citenamefont{Gelbwaser-Klimovsky
  et~al.}(2013)\citenamefont{Gelbwaser-Klimovsky, Alicki, and
  Kurizki}}]{gelbwaser2013work}
\bibinfo{author}{\bibfnamefont{D.}~\bibnamefont{Gelbwaser-Klimovsky}},
  \bibinfo{author}{\bibfnamefont{R.}~\bibnamefont{Alicki}}, \bibnamefont{and}
  \bibinfo{author}{\bibfnamefont{G.}~\bibnamefont{Kurizki}},
  \bibinfo{journal}{EPL} \textbf{\bibinfo{volume}{103}}, \bibinfo{pages}{60005}
  (\bibinfo{year}{2013}).

\bibitem[{\citenamefont{Gelbwaser-Klimovsky and
  Kurizki}(2014)}]{gelbwaser2014heat}
\bibinfo{author}{\bibfnamefont{D.}~\bibnamefont{Gelbwaser-Klimovsky}}
  \bibnamefont{and} \bibinfo{author}{\bibfnamefont{G.}~\bibnamefont{Kurizki}},
  \bibinfo{journal}{Physical Review E} \textbf{\bibinfo{volume}{90}},
  \bibinfo{pages}{022102} (\bibinfo{year}{2014}).

\bibitem[{\citenamefont{Mohseni et~al.}(2008)\citenamefont{Mohseni, Rebentrost,
  Lloyd, and Aspuru-Guzik}}]{mohseni2008environment}
\bibinfo{author}{\bibfnamefont{M.}~\bibnamefont{Mohseni}},
  \bibinfo{author}{\bibfnamefont{P.}~\bibnamefont{Rebentrost}},
  \bibinfo{author}{\bibfnamefont{S.}~\bibnamefont{Lloyd}}, \bibnamefont{and}
  \bibinfo{author}{\bibfnamefont{A.}~\bibnamefont{Aspuru-Guzik}},
  \bibinfo{journal}{The Journal of chemical physics}
  \textbf{\bibinfo{volume}{129}}, \bibinfo{pages}{174106}
  (\bibinfo{year}{2008}).

\bibitem[{\citenamefont{Rebentrost et~al.}(2009)\citenamefont{Rebentrost,
  Mohseni, Kassal, Lloyd, and Aspuru-Guzik}}]{rebentrost2009environment}
\bibinfo{author}{\bibfnamefont{P.}~\bibnamefont{Rebentrost}},
  \bibinfo{author}{\bibfnamefont{M.}~\bibnamefont{Mohseni}},
  \bibinfo{author}{\bibfnamefont{I.}~\bibnamefont{Kassal}},
  \bibinfo{author}{\bibfnamefont{S.}~\bibnamefont{Lloyd}}, \bibnamefont{and}
  \bibinfo{author}{\bibfnamefont{A.}~\bibnamefont{Aspuru-Guzik}},
  \bibinfo{journal}{New Journal of Physics} \textbf{\bibinfo{volume}{11}},
  \bibinfo{pages}{033003} (\bibinfo{year}{2009}).

\bibitem[{\citenamefont{Plenio and Huelga}(2008)}]{plenio2008dephasing}
\bibinfo{author}{\bibfnamefont{M.~B.} \bibnamefont{Plenio}} \bibnamefont{and}
  \bibinfo{author}{\bibfnamefont{S.~F.} \bibnamefont{Huelga}},
  \bibinfo{journal}{New Journal of Physics} \textbf{\bibinfo{volume}{10}},
  \bibinfo{pages}{113019} (\bibinfo{year}{2008}).

\bibitem[{\citenamefont{Novoderezhkin et~al.}(2004)\citenamefont{Novoderezhkin,
  Yakovlev, Van~Grondelle, and Shuvalov}}]{novoderezhkin2004coherent}
\bibinfo{author}{\bibfnamefont{V.~I.} \bibnamefont{Novoderezhkin}},
  \bibinfo{author}{\bibfnamefont{A.~G.} \bibnamefont{Yakovlev}},
  \bibinfo{author}{\bibfnamefont{R.}~\bibnamefont{Van~Grondelle}},
  \bibnamefont{and} \bibinfo{author}{\bibfnamefont{V.~A.}
  \bibnamefont{Shuvalov}}, \bibinfo{journal}{The Journal of Physical Chemistry
  B} \textbf{\bibinfo{volume}{108}}, \bibinfo{pages}{7445}
  (\bibinfo{year}{2004}).

\bibitem[{\citenamefont{Killoran et~al.}(2014)\citenamefont{Killoran, Huelga,
  and Plenio}}]{killoran2014enhancing}
\bibinfo{author}{\bibfnamefont{N.}~\bibnamefont{Killoran}},
  \bibinfo{author}{\bibfnamefont{S.~F.} \bibnamefont{Huelga}},
  \bibnamefont{and} \bibinfo{author}{\bibfnamefont{M.~B.}
  \bibnamefont{Plenio}}, \bibinfo{journal}{arXiv preprint arXiv:1412.4136}
  (\bibinfo{year}{2014}).

\bibitem[{\citenamefont{Dorfman et~al.}(2013)\citenamefont{Dorfman, Voronine,
  Mukamel, and Scully}}]{dorfman2013photosynthetic}
\bibinfo{author}{\bibfnamefont{K.~E.} \bibnamefont{Dorfman}},
  \bibinfo{author}{\bibfnamefont{D.~V.} \bibnamefont{Voronine}},
  \bibinfo{author}{\bibfnamefont{S.}~\bibnamefont{Mukamel}}, \bibnamefont{and}
  \bibinfo{author}{\bibfnamefont{M.~O.} \bibnamefont{Scully}},
  \bibinfo{journal}{Proceedings of the National Academy of Sciences}
  \textbf{\bibinfo{volume}{110}}, \bibinfo{pages}{2746} (\bibinfo{year}{2013}).

\bibitem[{\citenamefont{Scully et~al.}(2011)\citenamefont{Scully, Chapin,
  Dorfman, Kim, and Svidzinsky}}]{scully2011quantum}
\bibinfo{author}{\bibfnamefont{M.~O.} \bibnamefont{Scully}},
  \bibinfo{author}{\bibfnamefont{K.~R.} \bibnamefont{Chapin}},
  \bibinfo{author}{\bibfnamefont{K.~E.} \bibnamefont{Dorfman}},
  \bibinfo{author}{\bibfnamefont{M.~B.} \bibnamefont{Kim}}, \bibnamefont{and}
  \bibinfo{author}{\bibfnamefont{A.}~\bibnamefont{Svidzinsky}},
  \bibinfo{journal}{Proceedings of the National Academy of Sciences}
  \textbf{\bibinfo{volume}{108}}, \bibinfo{pages}{15097}
  (\bibinfo{year}{2011}).

\bibitem[{\citenamefont{Creatore et~al.}(2013)\citenamefont{Creatore, Parker,
  Emmott, and Chin}}]{creatore2013efficient}
\bibinfo{author}{\bibfnamefont{C.}~\bibnamefont{Creatore}},
  \bibinfo{author}{\bibfnamefont{M.}~\bibnamefont{Parker}},
  \bibinfo{author}{\bibfnamefont{S.}~\bibnamefont{Emmott}}, \bibnamefont{and}
  \bibinfo{author}{\bibfnamefont{A.}~\bibnamefont{Chin}},
  \bibinfo{journal}{Physical review letters} \textbf{\bibinfo{volume}{111}},
  \bibinfo{pages}{253601} (\bibinfo{year}{2013}).

\bibitem[{\citenamefont{Fassioli et~al.}(2010)\citenamefont{Fassioli, Nazir,
  and Olaya-Castro}}]{fassioli2010quantum}
\bibinfo{author}{\bibfnamefont{F.}~\bibnamefont{Fassioli}},
  \bibinfo{author}{\bibfnamefont{A.}~\bibnamefont{Nazir}}, \bibnamefont{and}
  \bibinfo{author}{\bibfnamefont{A.}~\bibnamefont{Olaya-Castro}},
  \bibinfo{journal}{The Journal of Physical Chemistry Letters}
  \textbf{\bibinfo{volume}{1}}, \bibinfo{pages}{2139} (\bibinfo{year}{2010}).

\bibitem[{\citenamefont{Alharbi and Kais}(2015)}]{alharbi2015theoretical}
\bibinfo{author}{\bibfnamefont{F.~H.} \bibnamefont{Alharbi}} \bibnamefont{and}
  \bibinfo{author}{\bibfnamefont{S.}~\bibnamefont{Kais}},
  \bibinfo{journal}{Renewable and Sustainable Energy Reviews}
  \textbf{\bibinfo{volume}{43}}, \bibinfo{pages}{1073} (\bibinfo{year}{2015}).

\bibitem[{\citenamefont{Cao and Silbey}(2009)}]{cao2009optimization}
\bibinfo{author}{\bibfnamefont{J.}~\bibnamefont{Cao}} \bibnamefont{and}
  \bibinfo{author}{\bibfnamefont{R.~J.} \bibnamefont{Silbey}},
  \bibinfo{journal}{J. Phys. Chem. A} \textbf{\bibinfo{volume}{113}},
  \bibinfo{pages}{13825} (\bibinfo{year}{2009}).

\bibitem[{\citenamefont{Spohn}(1978)}]{spohn1978entropy}
\bibinfo{author}{\bibfnamefont{H.}~\bibnamefont{Spohn}},
  \bibinfo{journal}{Journal of Mathematical Physics}
  \textbf{\bibinfo{volume}{19}}, \bibinfo{pages}{1227} (\bibinfo{year}{1978}).

\bibitem[{\citenamefont{Breuer and Petruccione}(2002)}]{breuer2002theory}
\bibinfo{author}{\bibfnamefont{H.-P.} \bibnamefont{Breuer}} \bibnamefont{and}
  \bibinfo{author}{\bibfnamefont{F.}~\bibnamefont{Petruccione}},
  \emph{\bibinfo{title}{The theory of open quantum systems}}
  (\bibinfo{publisher}{Oxford university press}, \bibinfo{year}{2002}).

\bibitem[{\citenamefont{Caruso et~al.}(2009)\citenamefont{Caruso, Chin, Datta,
  Huelga, and Plenio}}]{caruso2009highly}
\bibinfo{author}{\bibfnamefont{F.}~\bibnamefont{Caruso}},
  \bibinfo{author}{\bibfnamefont{A.~W.} \bibnamefont{Chin}},
  \bibinfo{author}{\bibfnamefont{A.}~\bibnamefont{Datta}},
  \bibinfo{author}{\bibfnamefont{S.~F.} \bibnamefont{Huelga}},
  \bibnamefont{and} \bibinfo{author}{\bibfnamefont{M.~B.}
  \bibnamefont{Plenio}}, \bibinfo{journal}{The Journal of Chemical Physics}
  \textbf{\bibinfo{volume}{131}}, \bibinfo{pages}{105106}
  (\bibinfo{year}{2009}).

\bibitem[{\citenamefont{Davies}(1974)}]{davies1974markovian}
\bibinfo{author}{\bibfnamefont{E.~B.} \bibnamefont{Davies}},
  \bibinfo{journal}{Communications in mathematical Physics}
  \textbf{\bibinfo{volume}{39}}, \bibinfo{pages}{91} (\bibinfo{year}{1974}).

\bibitem[{\citenamefont{Gorini et~al.}(1976)\citenamefont{Gorini, Kossakowski,
  and Sudarshan}}]{gorini1976completely}
\bibinfo{author}{\bibfnamefont{V.}~\bibnamefont{Gorini}},
  \bibinfo{author}{\bibfnamefont{A.}~\bibnamefont{Kossakowski}},
  \bibnamefont{and} \bibinfo{author}{\bibfnamefont{E.~C.~G.}
  \bibnamefont{Sudarshan}}, \bibinfo{journal}{Journal of Mathematical Physics}
  \textbf{\bibinfo{volume}{17}}, \bibinfo{pages}{821} (\bibinfo{year}{1976}).

\bibitem[{\citenamefont{Lindblad}(1976)}]{lindblad1976generators}
\bibinfo{author}{\bibfnamefont{G.}~\bibnamefont{Lindblad}},
  \bibinfo{journal}{Communications in Mathematical Physics}
  \textbf{\bibinfo{volume}{48}}, \bibinfo{pages}{119} (\bibinfo{year}{1976}).

\bibitem[{\citenamefont{Gordon et~al.}(2009)\citenamefont{Gordon, Bensky,
  Gelbwaser-Klimovsky, Rao, Erez, and Kurizki}}]{gordon2009cooling}
\bibinfo{author}{\bibfnamefont{G.}~\bibnamefont{Gordon}},
  \bibinfo{author}{\bibfnamefont{G.}~\bibnamefont{Bensky}},
  \bibinfo{author}{\bibfnamefont{D.}~\bibnamefont{Gelbwaser-Klimovsky}},
  \bibinfo{author}{\bibfnamefont{D.~B.} \bibnamefont{Rao}},
  \bibinfo{author}{\bibfnamefont{N.}~\bibnamefont{Erez}}, \bibnamefont{and}
  \bibinfo{author}{\bibfnamefont{G.}~\bibnamefont{Kurizki}},
  \bibinfo{journal}{New Journal of Physics} \textbf{\bibinfo{volume}{11}},
  \bibinfo{pages}{123025} (\bibinfo{year}{2009}).

\bibitem[{\citenamefont{Jones}(2009)}]{jones2009petite}
\bibinfo{author}{\bibfnamefont{M.}~\bibnamefont{Jones}},
  \bibinfo{journal}{Biochemical Society Transactions}
  \textbf{\bibinfo{volume}{37}}, \bibinfo{pages}{400} (\bibinfo{year}{2009}).

\bibitem[{\citenamefont{Holstein and Primakoff}(1940)}]{holstein1940field}
\bibinfo{author}{\bibfnamefont{T.}~\bibnamefont{Holstein}} \bibnamefont{and}
  \bibinfo{author}{\bibfnamefont{H.}~\bibnamefont{Primakoff}},
  \bibinfo{journal}{Physical Review} \textbf{\bibinfo{volume}{58}},
  \bibinfo{pages}{1098} (\bibinfo{year}{1940}).

\bibitem[{\citenamefont{Emary and Brandes}(2004)}]{emary2004phase}
\bibinfo{author}{\bibfnamefont{C.}~\bibnamefont{Emary}} \bibnamefont{and}
  \bibinfo{author}{\bibfnamefont{T.}~\bibnamefont{Brandes}},
  \bibinfo{journal}{Physical Review A} \textbf{\bibinfo{volume}{69}},
  \bibinfo{pages}{053804} (\bibinfo{year}{2004}).

\bibitem[{\citenamefont{Adolphs and Renger}(2006)}]{adolphs2006proteins}
\bibinfo{author}{\bibfnamefont{J.}~\bibnamefont{Adolphs}} \bibnamefont{and}
  \bibinfo{author}{\bibfnamefont{T.}~\bibnamefont{Renger}},
  \bibinfo{journal}{Biophysical journal} \textbf{\bibinfo{volume}{91}},
  \bibinfo{pages}{2778} (\bibinfo{year}{2006}).

\bibitem[{\citenamefont{Valleau et~al.}(2014)\citenamefont{Valleau, Saikin,
  Ansari-Oghol-Beig, Rostami, Mossallaei, and
  Aspuru-Guzik}}]{valleau2014electromagnetic}
\bibinfo{author}{\bibfnamefont{S.}~\bibnamefont{Valleau}},
  \bibinfo{author}{\bibfnamefont{S.~K.} \bibnamefont{Saikin}},
  \bibinfo{author}{\bibfnamefont{D.}~\bibnamefont{Ansari-Oghol-Beig}},
  \bibinfo{author}{\bibfnamefont{M.}~\bibnamefont{Rostami}},
  \bibinfo{author}{\bibfnamefont{H.}~\bibnamefont{Mossallaei}},
  \bibnamefont{and}
  \bibinfo{author}{\bibfnamefont{A.}~\bibnamefont{Aspuru-Guzik}},
  \bibinfo{journal}{ACS nano} \textbf{\bibinfo{volume}{8}},
  \bibinfo{pages}{3884} (\bibinfo{year}{2014}).

\bibitem[{\citenamefont{Van~Kampen}(1992)}]{van1992stochastic}
\bibinfo{author}{\bibfnamefont{N.~G.} \bibnamefont{Van~Kampen}},
  \emph{\bibinfo{title}{Stochastic processes in physics and chemistry}},
  vol.~\bibinfo{volume}{1} (\bibinfo{publisher}{Elsevier},
  \bibinfo{year}{1992}).

\end{thebibliography}

\end{document}